\documentclass[10pt,journal,compsoc]{IEEEtran}

\usepackage{booktabs}
\usepackage{graphicx}
\usepackage{amsmath}

\usepackage{algorithm}
\usepackage{algpseudocode}
\usepackage{algorithmicx}

\usepackage{array}
\usepackage{float}
\usepackage{url}
\usepackage{colortbl}
\usepackage{enumitem}
\usepackage{makecell}
\usepackage{multicol}
\usepackage{multirow}
\usepackage{slashbox}
\usepackage{amsmath}
\usepackage{amssymb}

\usepackage{caption}
\usepackage{subcaption}
\usepackage{balance}

\newtheorem{definition}{Definition}

\ifCLASSOPTIONcompsoc
  \usepackage[nocompress]{cite}
\else
  \usepackage{cite}
\fi

\begin{document}
\title{\textit{MeetSense}: A Lightweight Framework for Group Identification using Smartphones}

\author{Snigdha Das,
	Soumyajit Chatterjee,
	Sandip Chakraborty,
	Bivas Mitra % <-this % stops a space
	\IEEEcompsocitemizethanks{\IEEEcompsocthanksitem S. Das, S. Chatterjee, S. Chakraborty and B. Mitra are with the Department of Computer Science and Engineering, Indian Institute of Technology Kharagpur, India.\protect\\
		E-mail: snigdhadas@sit.iitkgp.ernet.in, sjituit@gmail.com, sandipc@cse.iitkgp.ac.in, bivas@cse.iitkgp.ac.in}%
	%	\IEEEcompsocthanksitem S. Chatterjee, S. Chakraborty and B. Mitra are with Anonymous University.}% <-this % stops an unwanted space
	}
%\author{Michael~Shell,~\IEEEmembership{Member,~IEEE,}
%        John~Doe,~\IEEEmembership{Fellow,~OSA,}
%        and~Jane~Doe,~\IEEEmembership{Life~Fellow,~IEEE}% <-this % stops a space
%\IEEEcompsocitemizethanks{\IEEEcompsocthanksitem M. Shell was with the Department
%of Electrical and Computer Engineering, Georgia Institute of Technology, Atlanta,
%GA, 30332.\protect\\
%% note need leading \protect in front of \\ to get a newline within \thanks as
%% \\ is fragile and will error, could use \hfil\break instead.
%E-mail: see http://www.michaelshell.org/contact.html
%\IEEEcompsocthanksitem J. Doe and J. Doe are with Anonymous University.}% <-this % stops an unwanted space

\markboth{IEEE Transactions on Mobile Computing,~Vol.~14, No.~8, August~2015}%
{Das \MakeLowercase{\textit{et al.}}: \textit{MeetSense}: A Lightweight Framework for Group Identification using Smartphones}
\IEEEtitleabstractindextext{%
\begin{abstract}
In an organization, individuals prefer to form various formal and informal groups for mutual interactions. Therefore, ubiquitous identification of such groups and understanding their dynamics are important to monitor activities, behaviours and well-being of the individuals. In this paper, we develop a lightweight, yet near-accurate, methodology, called \textit{MeetSense}, to identify various interacting groups based on collective sensing through users' smartphones. Group detection from sensor signals is not straightforward because users in proximity may not always be under the same group. Therefore, we use acoustic context extracted from audio signals to infer interaction pattern among the subjects in proximity. We have developed an unsupervised and lightweight mechanism for user group detection by taking cues from network science and measuring the cohesivity of the detected groups in terms of modularity. Taking modularity into consideration, \textit{MeetSense} can efficiently eliminate incorrect groups, as well as adapt the mechanism depending on the role played by the proximity and the acoustic context in a specific scenario.  The proposed method has been implemented and tested under many real-life scenarios in an academic institute environment, and we observe that \textit{MeetSense} can identify user groups with close to $90\%$ accuracy even in a noisy environment. 
%We consider two modes of sensing -- (a) proximity of users, (b) ambience. 
%The detection of user groups involves two levels of sensing -- (a) exploring the proximity of various users and (b) understanding the environment based on interaction, like all the users are in the same meeting room. 
%In contrary to the existing literature, we use unsupervised learning based approach to figure out the users in proximity, based on WiFi signal strength sensed at the smartphones. However, proximity alone is not sufficient to capture groups, because two users in proximity may be within two neighbouring rooms or in a big room. Therefore, we also sense context through smartphone sound sensor to capture that users in proximity are also within the same environment, which gives the signature of an interacting group. The proposed method has been implemented and tested under many real-life scenarios in an academic institute environment, and we observe that \textit{MeetSense} can identify user groups with close to $90\%$ accuracy. On the other hand, we find that the system is significantly lightweight that can be implemented on smartphones or similar smart handheld devices. 
\end{abstract}

}

\maketitle

\IEEEdisplaynontitleabstractindextext
\IEEEpeerreviewmaketitle

\IEEEraisesectionheading{\section{Introduction}\label{intro}}
%\section{Introduction} \label{intro}

\begin{figure}[!t]
	\centering
	\captionsetup{justification=centering}
	\includegraphics[clip,width=\columnwidth,keepaspectratio]{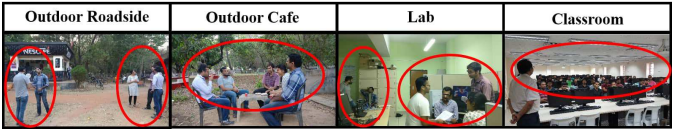}
	\caption{Setup of Different Meeting Group Scenarios}
	\label{fig:group}
\end{figure}

\IEEEPARstart{W}{orkplace} meetings and team formation among the individuals are key factors behind organizational efficiency. In organizations and institutions, people formally as well as sporadically meet, interact and form groups for various purposes, which include information sharing~\cite{mccomas2003citizen}, teaching and learning~\cite{clark1998teaching}, problem solving and decision making~\cite{mccomas2007predicting}, brainstorming~\cite{reinig2002dynamic}, socialization~\cite{horan2002effective} etc. Tracking the dynamics of group formation facilitates various utilities; for instance, organizational leaders may prefer to monitor the formation of teams, which benefit the overall efficiency and activeness of the organization~\cite{a2014understanding,jayarajah2015need}; course instructors in flipped classrooms~\cite{gilboy2015enhancing} in an academic environment may like to know how the students form groups among themselves to solve assignments and exercises. Unlike regular \& pre-scheduled team meetings, the formation of sporadic and instantaneous groups (often observed in office breaks, conferences etc.) make the problem challenging. On the other hand, increasing availability of sensor-rich smartphones provides a unique opportunity for collecting wide sensor information in a seamless manner. In this backdrop, we investigate the potential of smartphones to develop a lightweight ubiquitous system for identifying and monitoring group formation. Notably, in this paper, we primarily concentrate on the \emph{meeting} groups where co-located group members occasionally interact with each other. In this line, we capture different types of real-life meeting group scenarios such as outdoor roadside informal meeting; informal outdoor cafe meet, formal and informal laboratory meeting, and classroom interaction as shown in Figure \ref{fig:group}. 

%Tracking and monitoring such groups are important to understand the behaviour and the performance of individuals because the psychological states of individuals also depend on the surrounding environments and interactions among team members~\cite{jayarajah2015need}. The question that arises here is how we can develop a. Additionally, analysing such kind of groups require data for a long duration. It is pertinent to tell that for data analysing in such kind of applications, the real-time analysis is not much help.

Identification of a meeting group primarily relies on the location proximity~\cite{weppner2013bluetooth,sapiezynski2017} of the group members, which (apparently) can be conceptualized as a localization problem~\cite{wang2012no,wu2017gain}. In that direction, prior art explores the following three modalities -- GPS, Bluetooth, and WiFi for identification of the location similarity in supervised \cite{sapiezynski2017} as well as unsupervised \cite{casagranda2015audio} manner. In our context of group detection, vanilla localization based solutions demand high accuracy, which pushes the system towards complex processing. Notably, location proximity alone is insufficient to correctly discriminate and identify the meeting groups. For instance, consider a large conference hall, where multiple meeting groups get formed simultaneously; here members of different groups may exhibit location similarity among themselves, which makes the group detection challenging. Close inspection reveals that albeit similarity in location proximity, context~\cite{santani2015loud,calabrese2015urban,sen2014,azizyan2009surroundsense} of the members participating in individual meeting groups play a critical role in identifying groups; for instance all the members of a specific group in the conference hall share a substantial amount of contextual similarity (room illumination, ambience noise, member interactions, magnetic fluctuations)~\cite{sen2014,azizyan2009surroundsense,wang2012no}. However, identifying suitable contextual information, which is computationally lightweight as well as carries the signature of a meeting group, is an important problem. 

We propose \emph{acoustic context}, extracted from the audio signals received by individual smartphones, as a key context indicator. In order to compute \emph{acoustic context}, one can apply standard \textit{Mel-frequency Cepstral Coefficients} (MFCC)~\cite{davis1980comparison} on the recorded audio signals for speaker identification by measuring the tone \& pitch. However, this solution comes with multiple challenges. (a) The process of using MFCC usually follows a supervised approach which needs individuals' pre-trained information.
%MFCC is a supervised technique where the model needs to be individually pre-trained. 
%This severely reduces the potential of the model, as this restriction makes it unsuitable for instantaneous group detection, where most of the members are new and appear for the first time. 
(b) MFCC is computationally expensive, which makes it inappropriate for developing a lightweight system. 
%Moreover, the group detection problem does not compel us to exercise expensive operations such as speaker identification. 
(c) MFCC technique is quite sensitive to noise, hence most suitable for the unidirectional microphone with the stereo channel. Unfortunately, most of the commercial smartphones are equipped with omnidirectional microphones, which makes them prone to noise and corrupting the speaker identification process\footnote{\url{https://www.fabathome.org/best-smartphone-microphone/} (last accessed Apr 12, 2018) \label{footnote:microphone}}. In Next2Me~\cite{kar63969}, Baker and Efstratiou attempted to detect social groups considering WiFi and sound fingerprints. First, WiFi signal strengths are used for detecting the co-located population; next, this filtered population is fed to the audio module for finding out the social groups. The audio module considers the top $n$ frequencies of all the co-located individuals and computes the pairwise similarities. However, in the real-life environment, getting the actual top $n$ frequencies is challenging, and little variation in the selection of frequencies exerts a huge impact on the similarity computation.
Additionally, the audio signals captured on different smartphones can be time drifted, even if a single speaker acts as the audio source, since the clocks of different devices may not be time synchronized, and the subjects (devices) may be at different distances from the speaker. Once the co-located population has been identified and audio based context information has been extracted, state of the art techniques perform naive component analysis~\cite{kar63969} and community detection~\cite{pons2005computing} to identify social groups. However, in most of the cases, quality (cohesivity) of the discovered groups have been overlooked, which leads to the detection of incorrect communities (false positives). %In this paper, we take an important step to alleviate the aforesaid problems.

In this paper, we develop \textit{MeetSense}, a smartphone-driven ubiquitous platform for automatic detection of meeting groups. The proposed method is lightweight, unsupervised, hence equipped to detect instantaneously formed groups, without any pre-training. First, we determine the co-located population using standard localization techniques~\cite{sapiezynski2017,kar63969}. In our implementation, we relied on the WiFi-based proximity; nevertheless, this can be extended to Bluetooth and GPS based techniques as well. The crux of the proposed method is the computation of \emph{acoustic context} of the identified co-located population, which is based on the following key intuition. Interactions between participants of a meeting group switches from one speaker to another; where, at a time, there exists (mostly) one dominating speaker. Hence, power of the dominating tone (say $\alpha_1$) captured by the smartphones (subjects\footnote{In this paper, we use the term `subject' to indicate a participant, a member or a smartphone, interchangeably.}) in one group (say $G_6$) is significantly different from the power of the dominating tone ($\alpha_4$) captured by the devices of the another group $G_7$. If both the groups $G_6$ and $G_7$ are closely located, then all the devices might capture both the tones with varying power. However, for the devices in group $G_6$, the power of the dominating tone $\alpha_1$ should be higher than $\alpha_4$, whereas exactly opposite is likely to happen for the group $G_7$. By discriminating the power of the dominating tone, one can differentiate the acoustic context of the members of different groups. Finally, leveraging on the proximity of the co-located population and their acoustic context, we propose \textit{MeetSense}, a community-driven group detection model. The advantage of this model is manifold. 
\begin{enumerate}
	\item The model is unsupervised and lightweight. 
	\item This model can perform group detection even in the absence of proximity indicators (say, WiFi etc.). 
	\item We take cues from network science and measure the cohesivity of the detected groups with the help of \emph{modularity}. Taking modularity into consideration, \textit{MeetSense} can efficiently eliminate incorrect groups (reduce false positives), as well as adapt the algorithm depending on the role played by the proximity and the acoustic context in a specific scenario. For instance, in case of a noisy environment, \textit{MeetSense} combines both the modalities to identify meeting groups.
\end{enumerate}

The organization of the paper is the following. 
%In section~\ref{rwo} we briefly summarize the literature relevant to this work. 
In Section~\ref{so}, we formally define the meeting group and state the problem of group detection. We introduce two primary indicators and the related literature in those contexts -- (a) proximity, to identify co-located population and (b) audio signal, to compute acoustic context. We also conduct pilot experiments to highlight the challenges in extracting acoustic context amidst noisy environment, device heterogeneity etc.
%Then we focus on two signatures of the meeting groups based on a pilot study (\S\ref{so}) -- (a) proximity and (b) context.
%We develop a novel device independent, unsupervised approach for inferring users in proximity by exploiting WiFi signals captured by the smartphones.
In Section \ref{fim}, we propose a novel sound signal processing approach that can capture the acoustic context even with low power microphones available with the smartphones. In Section~\ref{gdm}, we develop \textit{MeetSense}, a group detection model leveraging on the community detection algorithms. We have implemented \textit{MeetSense} in an academic campus scenario, and captured several groups like classroom teaching, lab meetings, seminars, cafeteria gatherings, outdoor meetings etc. In Section~\ref{ds}, we show that \textit{MeetSense} can detect such groups with more than $90\%$ accuracy while incurring low computation overhead compared to the state of the art group identification methods.

\section{Problem Definition and Background Study} \label{so}
In this section, first, we define the meeting group and state the problem of group detection in the context of smartphone-based sensing. Next, we identify the primary indicators (say, proximity, acoustic context etc.) facilitating the group detection and explore their potential in the light of state of the art endeavours. Finally, we concentrate on the acoustic context and conduct a pilot study to highlight the challenges in group detection from audio signatures.
\subsection{Problem Statement}
We start with the definition of a \textit{Meeting Group} and subsequently state the problem of group detection.
\begin{definition}[Meeting Group]\label{def:group_new}
	Given a population of subjects $\mathbb{U}$, we define a meeting group $\mathcal{G}^{[t,t+T]} \subseteq \mathbb{U}$ for the time period $[t,t+T]$ as the collection of co-located individuals $\{u_i \in \mathbb{U}\}$ sharing similar context.
\end{definition}
For instance, two subjects $u_i$ and $u_j$ participate in a group $\mathcal{G}^{[t,t+T]}$ iff $u_i$ and $u_j$ are located in close proximity and share similar context for time duration $[t,t+T]$ \cite{casagranda2015audio}. In this paper, we hypothesize that sound profile, observed by the group members, defines the \emph{acoustic context} of a group. For instance, sensing the verbal interactions between the participating members can discriminate one meeting group from another. Notably, in the acoustic context, we only concentrate on the \emph{tone} from the human voice signal, and categorically disregard the content of interaction to preserve privacy.

Consider each subject $u_i \in \mathbb{U}$ carries a smartphone equipped with various sensors. We collect the sensor log $s_i$ from each subject $u_i$ and populate the data in a central repository $\mathcal{X}$. The sensor log $s_i$ comprises of the location information $p_i$ and acoustic information $\alpha_i$. The location information may come from various signals for indoor and outdoor localization techniques based on GPS, wireless signals etc.~\cite{chintalapudi2010indoor,azizyan2009surroundsense,abdelnasser2016semanticslam,wu2017gain,aly2017accurate,paek2010energy}; similarly acoustic information can be extracted from the audio signals captured by the smartphones~\cite{do2011groupus,hong2016socialprobe,lee2013sociophone,zhang2015tracking}. We aim to discover the meeting group $\mathcal{G}^{[t,t+T]}$ formed during the period $[t, t+T]$ from the logged sensor repository $\mathcal{X}$.

\subsection{Primary Indicators and Respective Prior Art}
The definition of the meeting group mainly relies on two kinds of sensing modalities --
(a) \textit{location Proximity} and (b) \textit{Acoustic context}. We explore the recent attempts in this direction and highlight their potential \& challenges in group detection.

%[BM: rewrite]Our further investigation on the origin of these modalities includes several aspects of smartphone-based sensing.
%Our further investigation on the origin of these modalities in smartphone-based tracking lands to two sensors -- (a) WiFi and (b) Microphone.

\subsubsection{Location Proximity} Localizing the subjects within their proximity is the initial step towards the group identification. In the line, the past literature explores mainly three modalities -- GPS, Bluetooth, and WiFi. GPS~\cite{aly2017accurate} is an important modality (albeit energy-hungry) for localization and detecting population within proximity. Although GPS performs well in outdoor environments, its accuracy sharply falls in indoor environments due to the interruption in the signal~\cite{chintalapudi2010indoor}. On the other side, Bluetooth-based study \cite{eagle2006reality} is one of the earliest attempts for localization in indoor environments. However, Bluetooth scanning is power hungry~\cite{sapiezynski2017}. Moreover, many of the Android smartphones (starting from versions $4.4$) have partial support for Bluetooth Low-Energy (BLE), which are capable of only detecting other BLE devices~\cite{sen2014}. Additionally, the Bluetooth signal as a medium of information is considered to be unreliable and noisy.

Recently, attempts have been made to detect proximity from WiFi fingerprint \cite{sapiezynski2017}. WiFi-based localization is considered as a promising indicator for identifying the population in proximity. WiFi consumes significantly less power as compared to Bluetooth and GPS. Although BLE appears as an alternative to WiFi in terms of power consumption, nevertheless BLE suffers from data loss and fluctuations with increasing distance~\cite{friedman2013power}. Furthermore, WiFi can work in any environment irrespective of whether the device location is indoor or outdoor. Each modality has its positive and negative aspects in the context of localization. Hence, the selection of modalities is highly dependent on the application for which the proximity is computed.  In~\cite{sapiezynski2017}, the authors have developed a supervised based learning approach for person-to-person proximity detection using WiFi fingerprints, like access point (AP) coverage and signal strength measurements. On the other hand, the authors in~\cite{kar63969} have developed an unsupervised learning based approach for proximity detection using a novel WiFi based metric computed using \textit{Manhattan distance}, which is the average of the pairwise signal strength difference among the APs, from which the subject receives signals. Any of these existing mechanism can be used for proximity detection for group identification. Once a set of subjects are detected to be in proximity, then the contextual similarity further characterizes the subset that forms a meeting group.

\begin{figure}[!t]
	\centering
	\captionsetup{justification=centering}
	\includegraphics[clip,height=80pt,keepaspectratio]{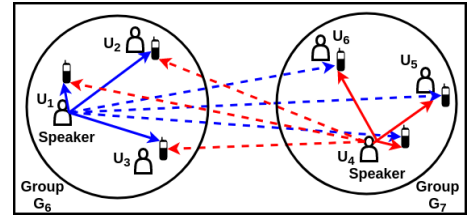}
	\caption{Impact of Audio Signals in Group Detection - two speakers from two different groups talk simultaneously}
	\label{fig:audioGroup}
\end{figure}

\subsubsection{Acoustic Context}
A microphone is an important indicator to identify the meeting group members. Participants, in general, avoid talking simultaneously in a meeting; although there can be a small overlap when the discussion switches from one speaker to another (utterance duration). Therefore the voice properties, such as pitch and tone of the current speaker in a group dominates in the audio signals captured by individual subjects in that group~\cite{microsoft1}. \textit{Pitch} defines the perceived fundamental frequency of the sound~\cite{yin}, whereas \textit{tone} is the variation or thickness of the pitch, indicating the quality of the sound~\cite{crowd}. Figure~\ref{fig:audioGroup} explains the intuition behind using human voice characteristics for group identification. The blue audio signal dominates for the subjects of $G_6$, whereas the red signal dominates for the subjects of $G_7$. Therefore, human voice characteristics (aka acoustic context) may show a strong feature similarity, if the subjects belong to the same group.

Audio pitch and tone extraction from human voice signal is a well-studied problem in the literature~\cite{crowd,yin}. YIN~\cite{yin} is a simple time-domain pitch calculation algorithm which is used in many existing applications such as counting the crowd from human voice signals~\cite{crowd}. Although the pitch is a good indicator for speaker identification, however, pitch alone fails to differentiate the relative distance of the speakers from other subjects, since it only concentrates on the central frequency of the audio signal. Therefore, tone information needs to be extracted along with the pitch, and \textit{Mel-frequency Cepstral Coefficients} (MFCC) based techniques~\cite{davis1980comparison} with Gaussian Mixture Model (GMM)~\cite{song2016detecting} can be applied for this purpose. However, in smartphones, the use of unidirectional microphones with the stereo channel is rare. A smartphone may capture the voice signals from the subjects of the other nearby groups, apart from the primary speaker of its group, as shown in Figure~\ref{fig:audioGroup}.  Further, the environmental noise generated from the variety of external sources may collude the recorded audio signal. For instance, the humming noise generated from the ACs and other machines (indoor) and vehicles (outdoor) may collude the collected audio signals and make the group detection challenging. Additionally, for instantaneous group detection, there is no apriori knowledge of the group members' tone information. Therefore, group identification from MFCC based audio processing along with some supervised techniques may pose some additional challenges, although they work well for applications like crowd count~\cite{crowd}. In the following, we explore these challenges from the observations over a pilot experiment.
%One single dominating audio tone from the speaker of a group, captured across the smartphones of all the subjects, indicates towards a meeting group.
%However, the environmental noise generated from the variety of external sources may collude recorded the audio signal. For instance, the humming noise generated from the AC machines (indoor) and vehicles (outdoor) may collude the collected audio signals and make the group detection challenging. Moreover, in smartphones, the use of unidirectional microphones with the stereo channel is rare. This makes the standard voice data modelling techniques, such as the combination of Mel-Frequency Cepstral Coefficients (MFCC) with Gaussian Mixture Model (GMM)~\cite{song2016detecting} unsuitable in case of smartphones. Hence, detecting the interaction in such constraint environmental conditions and device setup is challenging. Also, the heterogeneity of devices has an adverse impact on the amplitude of the recorded audio signal as the automatic gain control circuits attached to the smartphone microphones varies across the devices~\cite{microsoft1,microsoft2}.

\subsection{Pilot Study: Revealing Challenges with Audio Signals}
We launched a pilot study to examine the potential of the acoustic context in identifying the meeting groups amidst challenging scenarios.
%As we mentioned earlier, proximity is well studied in this context. Hence, we continued with the different variant of the existing approaches and excluded the proximity discussion in the pilot study.
We developed an Android app for collecting the audio signal log from the smartphones for conducting the pilot experiment. We recruited six subjects in this experiment for two weeks, installed the app on their smartphones and instructed them to occasionally form pre-designed meeting groups (multiple times) for around $T \geq 15$ minutes. Subjects have been asked to record the group formation instances manually for validation. The detailed overview of the formed groups in this study is listed in Table \ref{table:pilot}. During the experiments, we have captured $16$ bit audio signal at $44.1$ kHz sampling rate from the in-built microphones of the smartphones. Notably, while forming those controlled groups, we pay special attention towards incorporating two fundamental challenges that may affect the group identification from audio signals -- (1) device heterogeneity and (2) noisy environments. To capture device heterogeneity, we have used smartphones from four different makes and models -- $2$ Moto X, $1$ Moto G 2nd Gen, $2$ OnePlus3, $1$ Samsung Note5. The noise environment can be summarized in the following generic scenarios.

\begin{table}[!t]
	\centering
	\caption{Pilot Study Minutiae}
	\label{table:pilot}
	\begin{tabular}{|l|l|l|l|}
		\hline
		\multicolumn{1}{|c|}{\textbf{Group ID}} & \multicolumn{1}{c|}{\textbf{Member IDs}} & \multicolumn{1}{c|}{\textbf{Location}} & \textbf{\begin{tabular}[c]{@{}c}Primary\\Speaker\end{tabular}} \\ \hline
		$G_1$                                  & $U_1$,$U_3$,$U_4$                       & SMR Lab  & $U_4$                                  \\ \hline
		$G_2$                                  & $U_2$,$U_5$,$U_6$                       & Class C-118   & $U_2$                            \\ \hline
		$G_3$                                  & $U_1$,$U_2$,$U_3$,$U_4$                 & Cafeteria         & $U_3$                  \\ \hline
		$G_4$                                  & $U_2$,$U_5$                             & SMR Lab                      & $U_5$            \\ \hline
		$G_5$                                  & $U_1$,$U_2$,$U_3$,$U_4$                 & Way to Cafeteria                & $U_4$                  \\ \hline
		$G_6$                                  & $U_1$,$U_2$,$U_3$                             & Outdoor Roadside                      & $U_1$            \\ \hline
		$G_7$                                  & $U_4$,$U_5$,$U_6$                 & Outdoor Roadside                & $U_4$                  \\ \hline
	\end{tabular}
\end{table}

\begin{figure*}[!t]
	\captionsetup{justification=centering}
	\begin{minipage}{0.3\linewidth}
		\centering
		\includegraphics[clip,height=120pt,keepaspectratio]{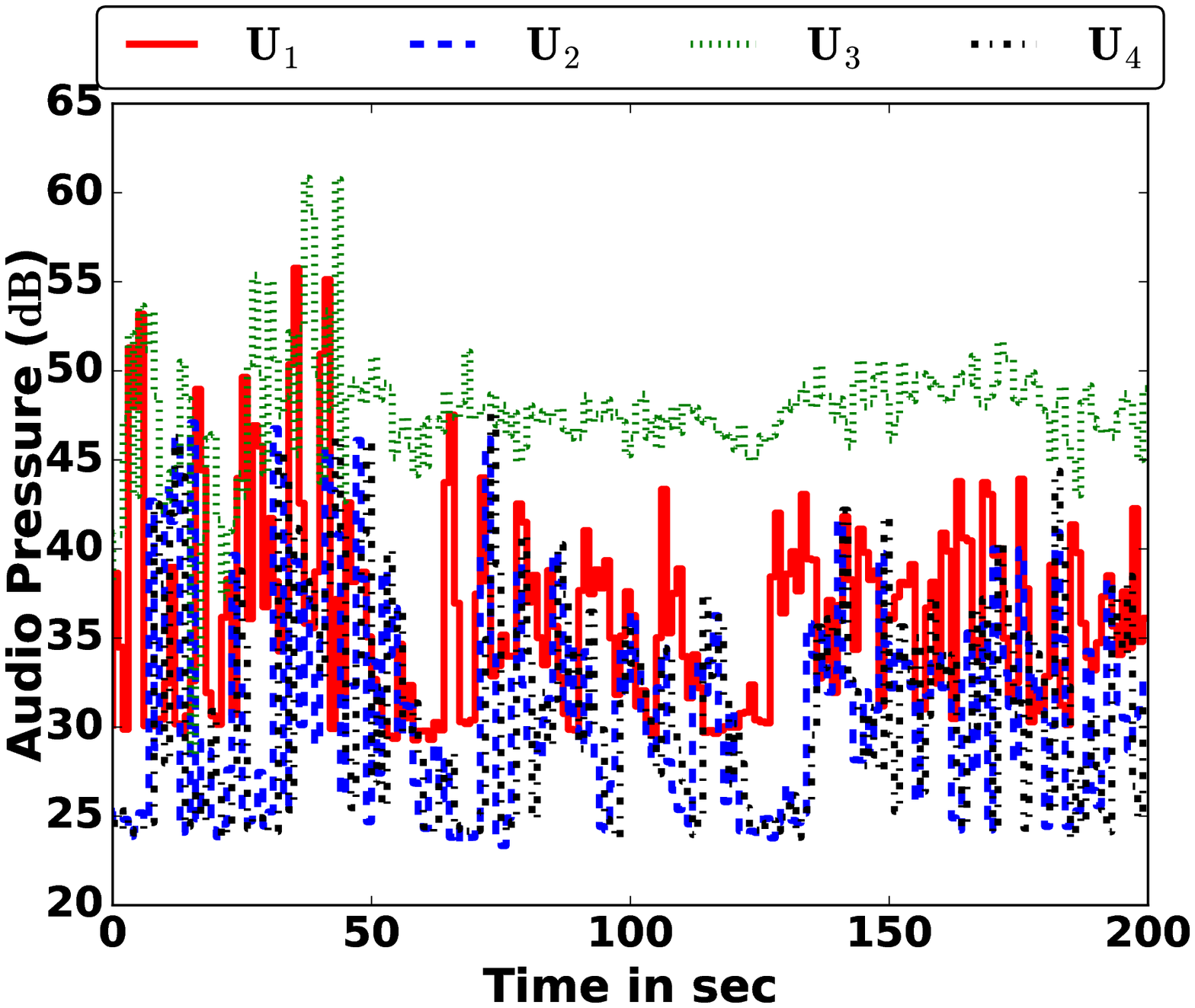}
		\centering \caption{Impact of Audio Pressure among the Subjects of Same Group: $U_1$, $U_2$, $U_3$, $U_4$ in $G_3$}
		\label{fig:audioVoice}
	\end{minipage}
	\hspace{0.1cm}
	\begin{minipage}{0.3\linewidth}
		\centering
		\includegraphics[clip,height=120pt,keepaspectratio]{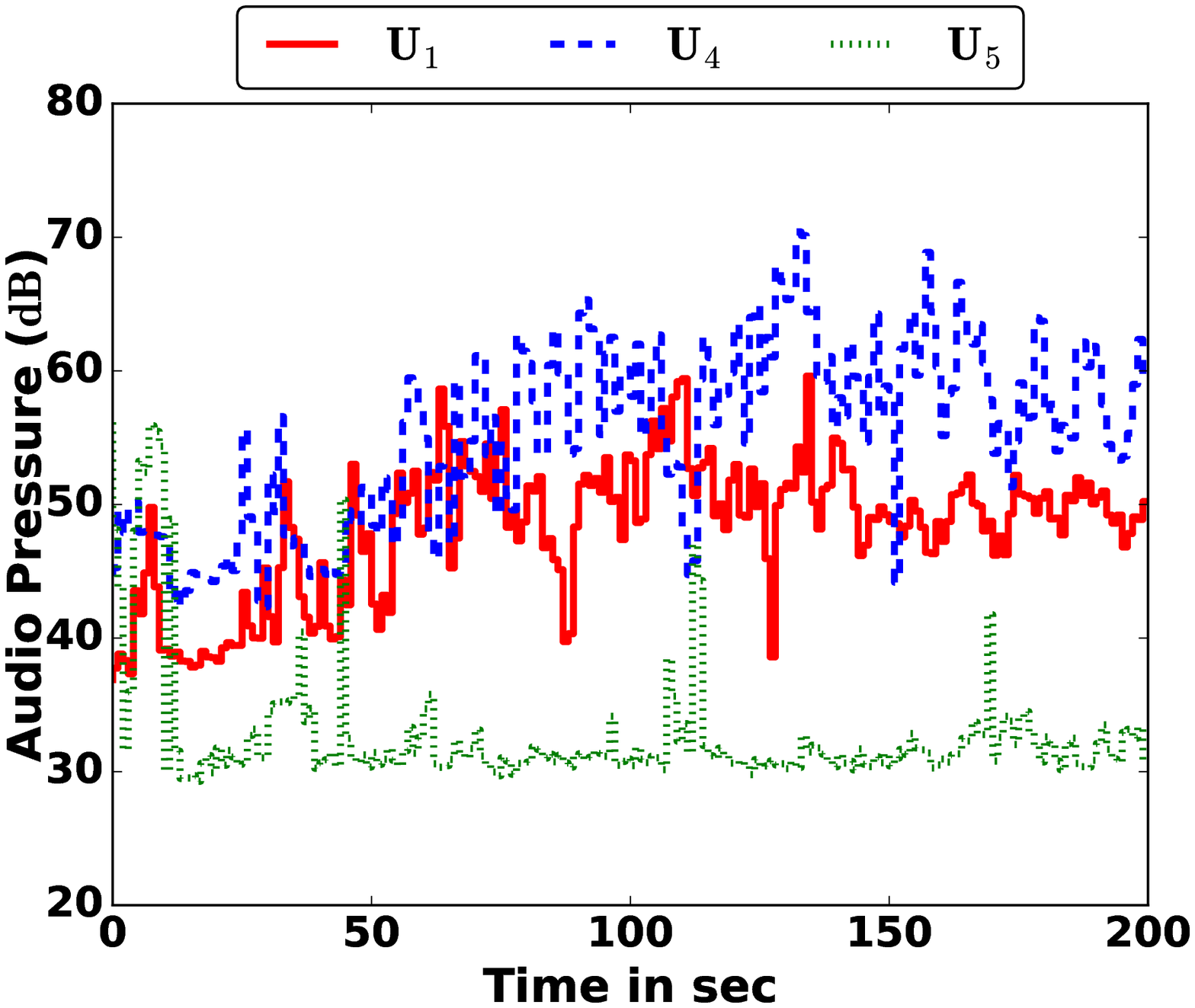}
		\centering \caption{Impact of Audio Pressure among the Subjects of Different Groups: $U_1$, $U_4$ in $G_1$ and $U_5$ in $G_4$}
		\label{fig:audioNoise}
	\end{minipage}
	\hspace{0.1cm}
	\begin{minipage}{0.35\linewidth}
		\centering
		\includegraphics[scale=0.2]{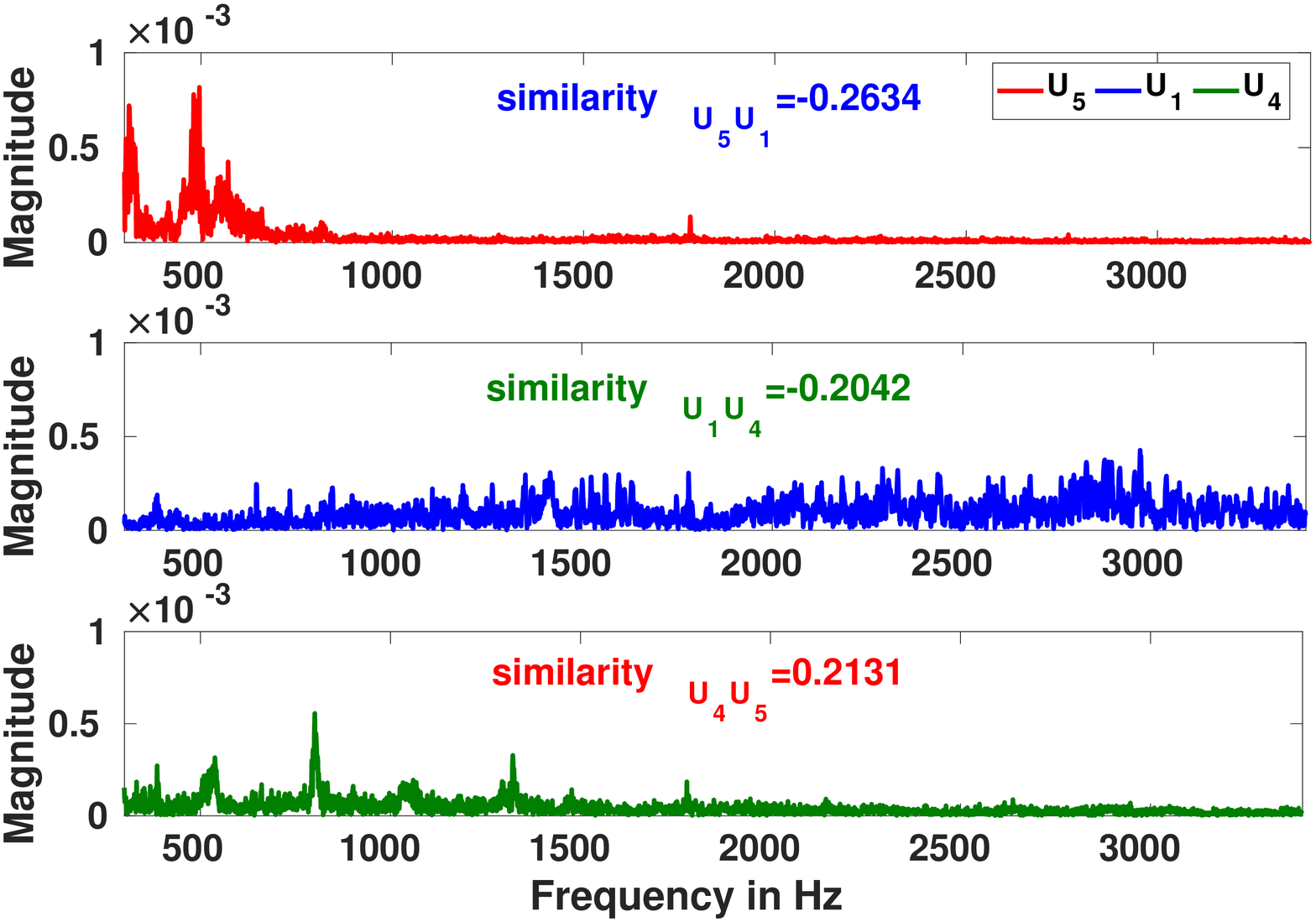}
		\caption{Deviations of Frequencies in Groups}
		\label{fig:aud_frq}
	\end{minipage}
\end{figure*}

\textbf{(a) Low noise environment:} This includes the formation of meeting groups where the surrounding environmental noise is low (audio amplitude less than $40$ dB \cite{audio-level}). Subjects forming groups inside classrooms (group $G_2$), during the conference in formal gatherings, while moving inside the laboratory ($G_1$, $G_4$) etc. can be included in this scenario.

\textbf{(b) Noisy environment:} In this scenario, subjects are forming groups in the highly noisy environment (audio amplitude more than $40$ dB). Subjects forming groups in cafeterias (group $G_3$), in informal gathering, marketplace, outdoor environment ($G_5$) etc. fall in this scenario.

\subsubsection{Observations}
%We turn our attention to the microphone where we capture the $16$ bit audio signal at $44.1$ kHz sampling rate.
We first normalize the amplitude of the audio signal and then compute the audio pressure as an indicator of the volume of the audio signal received by individual devices. We concentrate on the meeting group $G_3$ where subject $U_3$ primarily speaks while other group members mostly remain silent. In Figure \ref{fig:audioVoice}, we plot the audio pressure received from the individual subjects ($U_1$, $U_2$, $U_3$, $U_4$) of group $G_3$. We observe that subjects ($U_1$, $U_2$, $U_4$), participating in the same group $G_3$, exhibit similar audio pressure. However, the audio pressure of $U_3$ deviates from the rest of the subjects since the user is moving while speaking. Therefore, the values are slightly different than the other group members.

\begin{figure}[t]
	\centering
	\captionsetup{justification=centering}
	\begin{subfigure}[b]{0.24\textwidth}
		\centering
		\includegraphics[clip,height=80pt,keepaspectratio]{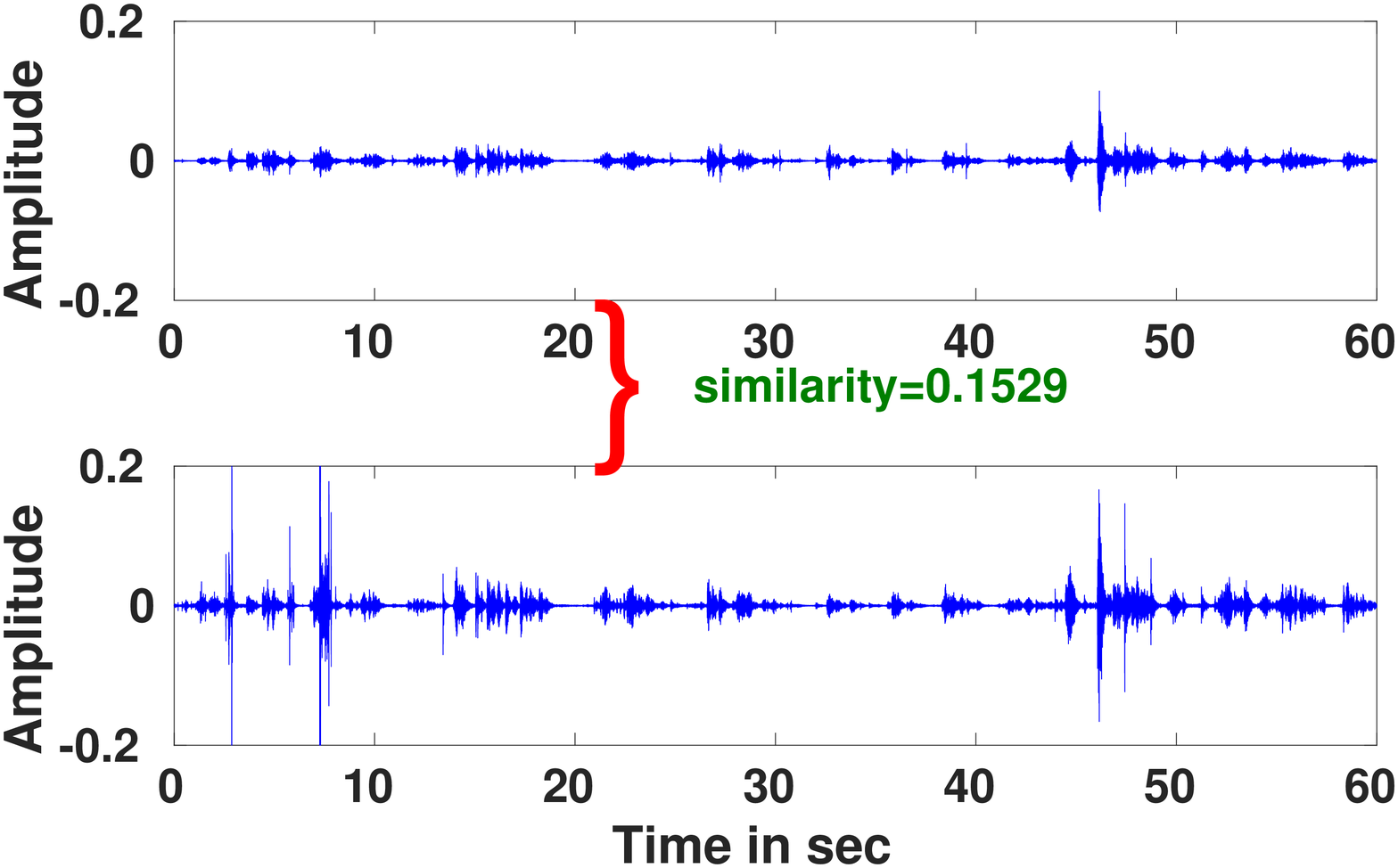}
		\caption{Same Build: \\$U_2$ and $U_5$ in $G_2$}
		\label{fig:audsame}
	\end{subfigure}%
	\hfill
	\begin{subfigure}[b]{0.24\textwidth}
		\centering
		\includegraphics[clip,height=80pt,keepaspectratio]{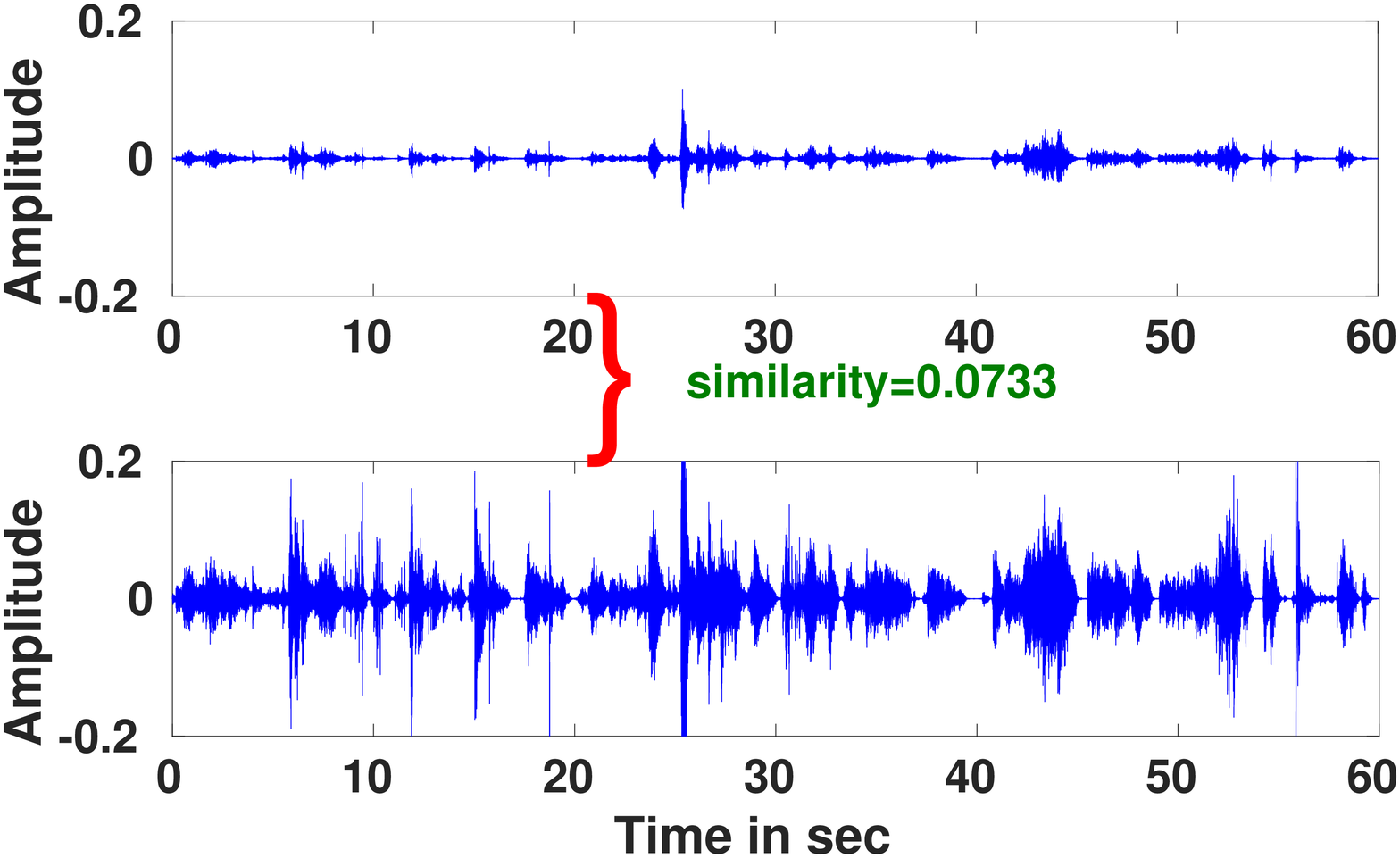}
		\caption{Different Build: \\$U_2$ and $U_6$ in $G_2$}
		\label{fig:auddiff}
	\end{subfigure}%
	\caption{Audio amplitude in devices of same and different builds}
	\label{fig:audio}
\end{figure}

The scenario gets compounded when we consider two groups $G_1$ ($U_1$, $U_3$, $U_4$) and $G_4$ ($U_2$, $U_5$) which get formed inside the same laboratory during the similar time period. Figure \ref{fig:audioNoise} highlights the fact that although audio pressure of the subjects ($U_1$, $U_4$) participating in same group ($G_1$) exhibit similar behavior, however, the same indicator fails to show clear discrimination between the subjects (say $U_1$, $U_5$) participating in two different groups ($G_1$, $G_4$). For further investigation, we move to frequency domain based analysis. Importantly, Figure \ref{fig:aud_frq} shows that the frequency component present in subject $U_1$ exhibits contrasting behaviour from the subject $U_5$, belonging to a different group. However, the frequency components of subjects $U_1$ and $U_4$ present in the same meeting group ($G_1$) exhibit (albeit minor) difference (due to environmental noise), posing a new challenge. Last but not the least, Figure \ref{fig:audio} demonstrates the variation of amplitude (raw version of audio pressure) due to device heterogeneity. The smartphone microphones use \textit{automatic gain control} (AGC) circuit, which exaggerates the variation of amplitude for the same audio signal captured through different devices. In group $G_2$, the subjects $U_2$ and $U_5$ carrying same make \& model devices whereas another subject $U_6$ carries a different build. Although all three of them belong to the same meeting group, Figure \ref{fig:auddiff} exhibits a dissimilarity in amplitude for the subjects $U_2$ and $U_6$ (nevertheless, similarity can be observed for subjects $U_2$ and $U_5$ (Figure~\ref{fig:audsame})). The detailed comparison of the devices is listed in Table \ref{aud_het}, in terms of similarity index for the same audio signal. We observe that the similarity index is sometimes quite low for two different devices from two different makes and models.

\subsubsection{Lessons Learnt}
We observe that audio signals provide us with a good indicator to capture the acoustic context of a group. However, due to omnidirectional nature of smartphone microphones, a significant audio pressure from the speakers of the nearby groups is also getting captured, as we observe in Figure \ref{fig:audioNoise} (formation of two groups $G_1$ and $G_4$ in the same lab). For MFCC based techniques, the separability of cepstral coefficients gets distorted in the presence of multiple speakers and environmental noise~\cite{mfcc_noise,microsoft1,microsoft2}. Hence, MFCC may be able to capture the presence of two speakers in the vicinity for groups $G_1$ and $G_4$, but it will fail to classify the subjects based on the primary speakers. Device heterogeneity exaggerates this problem further. In summary, although microphone provides important signature uncovering group membership, however, it is inadequate in its current form for handling adverse scenarios.

\begin{table}[!t]
	\centering
	\caption{Audio Amplitude Similarity in Heterogeneous Devices}
	\label{aud_het}
	\begin{tabular}{|c|l|l|l|l|}
		\hline
		\textbf{\begin{tabular}[c]{@{}c@{}}Device-Device\\ Audio Amplitude\\ Similarity\end{tabular}} & \multicolumn{1}{c|}{\textbf{MotoX}} & \multicolumn{1}{c|}{\textbf{\begin{tabular}[c]{@{}c@{}}Samsung\\ Note5\end{tabular}}} & \multicolumn{1}{c|}{\textbf{OnePlus3}} & \multicolumn{1}{c|}{\textbf{MotoG}} \\ \hline
		\textbf{Moto X}                                                                              & \textit{\textbf{0.3247}}             & 0.1178                                                                                & -0.2781                                & -0.1138                              \\ \hline
		\textbf{Samsung Note5}                                                                       & 0.1178                               & \textit{\textbf{0.2977}}                                                              & 0.0896                                 & 0.1287                               \\ \hline
		\textbf{Oneplus3}                                                                            & -0.2781                              & 0.0896                                                                                & \textit{\textbf{0.5671}}               & -0.0653                              \\ \hline
		\textbf{Moto G}                                                                              & -0.1138                              & 0.1287                                                                                & -0.0653                                & \textit{\textbf{0.5822}}                                   \\ \hline
	\end{tabular}
\end{table}

\section{Measuring Acoustic Context of Meeting Groups} \label{fim}
\begin{figure}[t]
	\centering
	\captionsetup{justification=centering}
	\begin{minipage}{.4\textwidth}
		\includegraphics[clip,height=64pt,keepaspectratio]{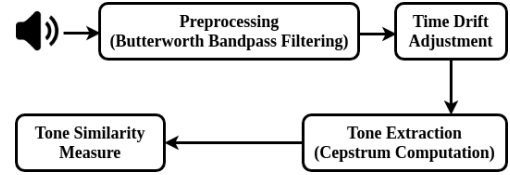}
		\caption{Audio Signal Processing Flowchart}
		\label{fig:audioFlow}
	\end{minipage}
\end{figure}

From the pilot study, we demonstrate that audio signals are rich sources to capture the context of a meeting group. However, we also comprehend that the naive audio processing techniques are not sufficient to extract reliable information under various complicated scenarios.
%The broad intuition is that in a meeting, people, in general, avoids talking simultaneously; although there can be a small overlap when the discussion switches from one speaker to another (utterance duration). Therefore the voice properties (such as pitch and tone) for the current speaker in a group dominates in the audio signals captured by individual subjects in that group~\cite{microsoft1}. Consider Figure~\ref{fig:audioGroup}. The blue audio signal is dominating for the subjects in Group 1, whereas the red signal is dominating for the subjects of Group 2. Therefore, if we extract human voice characteristics from the audio signals captured by two subjects and use them as the feature for interaction characterization, we can observe a strong feature similarity if both of them belong to the same group and a poor feature similarity if they belong to different groups.
In this section, we develop a methodology for computing acoustic context from smartphone audio signals, as shown in Figure~\ref{fig:audioFlow}. The different steps in this procedure are as follows.

\subsection{Preprocessing of Vanilla Audio Signals}
For audio-based feature extraction, we collect the audio data $\alpha_i$ from all the subjects $u_i$ at a sampling rate of $f_s$, continuously for $t$ units with an interval of $\hat{t}$ units of time, where $t$ and $\hat{t}$ are specified by the application developers. We first extract the human speech signal between $300$ Hz to $3400$ Hz using Butterworth bandpass filtering. The human speech signals captured from different smartphones are used for further processing.

\subsubsection{Time Drift Adjustment}

\begin{figure}[t]
	\centering
	\captionsetup{justification=centering}
	\begin{subfigure}[b]{0.24\textwidth}
		\includegraphics[clip,height=86pt,keepaspectratio]{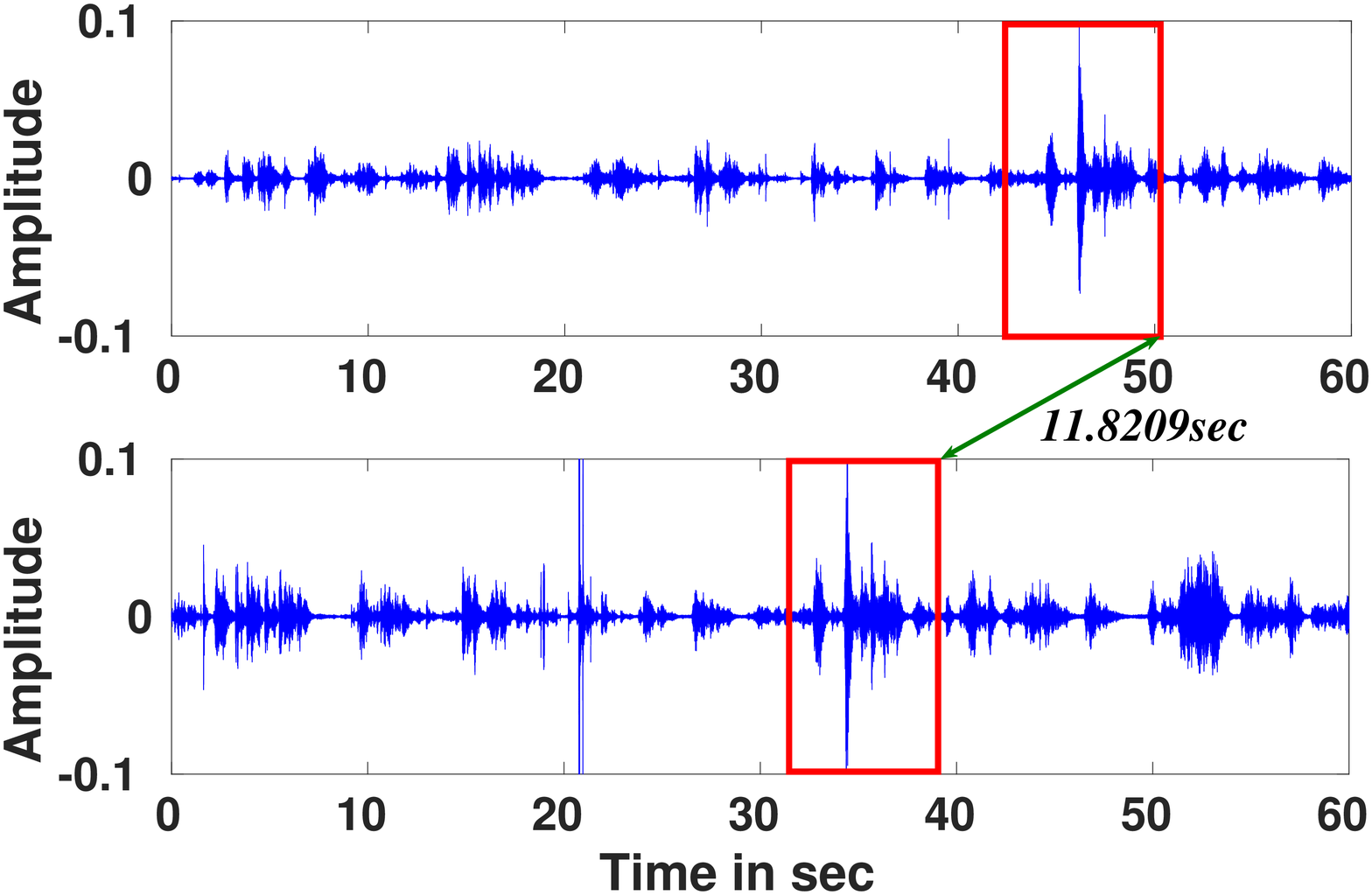}
		\caption{Two time drifted signals from same audio source}
		\label{fig:drift_y}
	\end{subfigure}
	\hfill
	\begin{subfigure}[b]{0.24\textwidth}
		\includegraphics[clip,height=80pt,keepaspectratio]{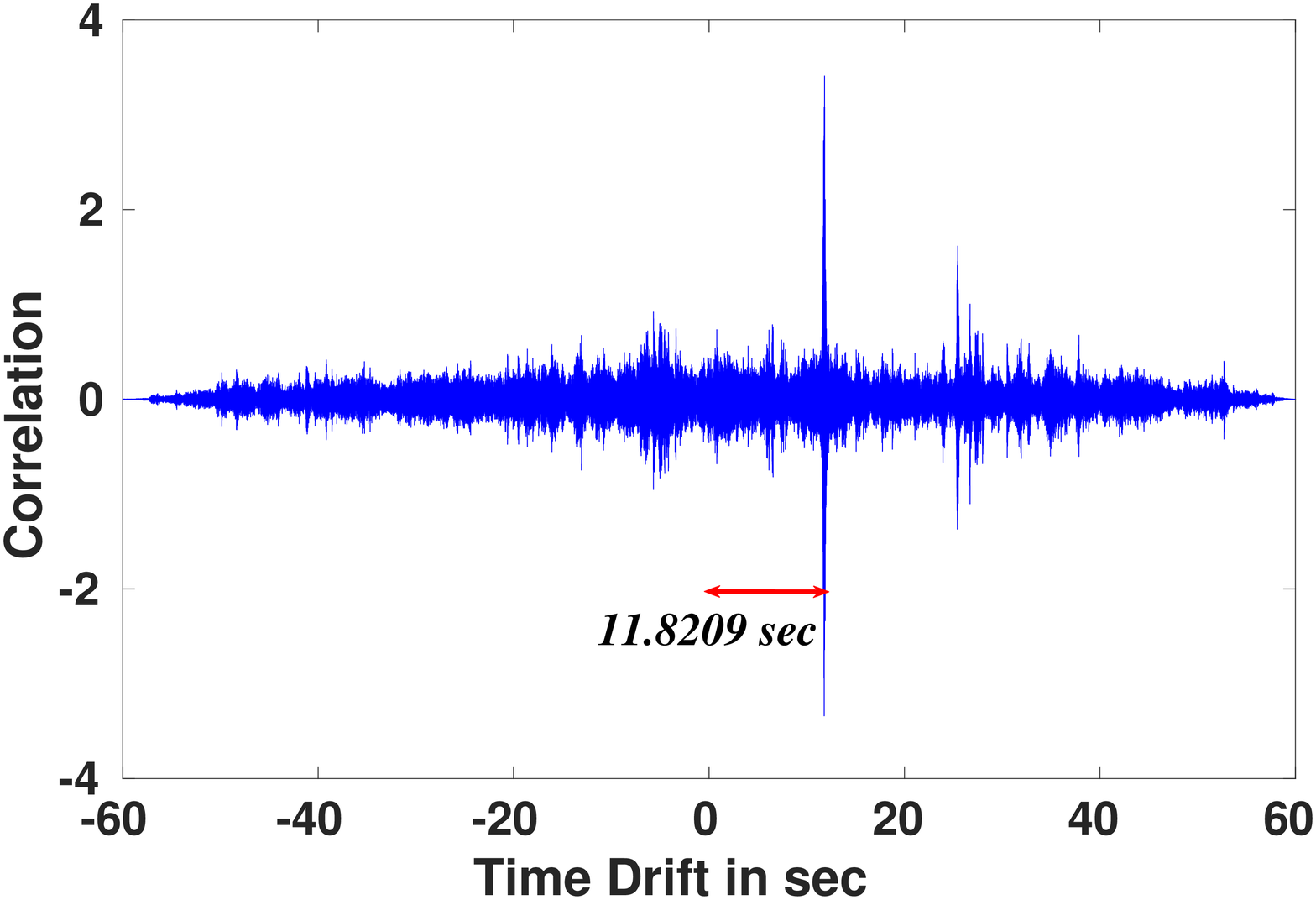}
		\caption{Correlation by shifting the second signal with reference to the first signal}
		\label{fig:drift_n}
	\end{subfigure}
	\caption{Computation of Time Drift}
	\label{fig:drift}
\end{figure}

The audio signals captured from different devices can be time drifted, even if a single speaker acts as the audio source. There are broadly two reasons for this -- (a) the clocks at different devices may not be time synchronized, and (b) the subjects may be at different distances from the speaker, which introduces propagation lag to the signals. Figure~\ref{fig:drift_y} shows the time drifted signals with a single speaker, captured from two different subjects. To compare two signals, we need to place both the signals at the same time reference frame, and therefore eliminating the time drift is an important task for audio processing.

Although some existing studies have developed techniques for time drift adjustment of audio signals captured in hand-held devices~\cite{time_drift1}, they employ smoothing techniques over the raw signal and thus tend to lose the physical properties of the signal, such as tone and pitch of the signal. However, such physical properties are important to capture the nature of human voice, which are essential for extracting acoustic context. Therefore, we apply a simple technique in this paper to mitigate the time drift introduced in the signals coming from a single audio source.

To eliminate the time drift, we use the concept of similarity measure between the signals in the time domain. Consider the audio signal coming from a single source, but captured at two different devices. Ideally, when both the signals are placed at the same reference frame, at the time domain (drift is zero), the similarity between them should be maximum. To measure the similarity between the signals, we use statistical correlation. The procedure works as follows. We fix one signal as the reference, and then shift another signal for the one-time unit at every step, and measure the correlation between the signals. Figure~\ref{fig:drift_n} plots the correlation between the signals shown in Figure~\ref{fig:drift_y}, with respect to the amount of time shift applied over the second signal, while considering the first signal as the reference. A positive time shift indicates that the second signal has been shifted towards the time clock, and a negative time shift represents that the signal has been shifted backwards the time clock. In the example, we observe that the correlation is maximum when the second signal is shifted $11.8209$ seconds, indicating that the drift is $11.8209$ seconds. Once the drift is calculated, one signal is shifted to make the drift zero with respect to the reference signal.

\subsection{Audio Tone Extraction}
The audio tone of the members of a meeting group should exhibit high similarity among themselves whereas tone dissimilarity indicates different groups. Hence, pairwise tone similarity between the group members should be an important property to determine the acoustic context of that group. Considering that group participants, in general, avoid talking simultaneously in a meeting, intuitively, there exists one dominating tone that gets captured at the smartphones of all the subjects in a meeting group. Audio tone extraction is a well-studied problem~\cite{crowd,yin} and \textit{Mel-frequency cepstral coefficients} (MFCC) based techniques~\cite{yin} are widely applied for tone extraction from audio signals. However, we face the following challenges while extracting the tone from smartphone audio signals. (a) Smartphone microphones are omnidirectional, and they capture environmental noise along with the human voice. Moreover, the devices are heterogeneous. MFCC fails in the face of the noisy environment and with device heterogeneity~\cite{mfcc_noise}. 
(b) The device heterogeneity is in general handled through various energy-based normalization techniques~\cite{crowd,microsoft1,microsoft2}, however they fail for smartphone microphones due to the nonlinearity gain of amplifiers and the presence of \textit{automatic gain control} (AGC) circuits\textsuperscript{\ref{footnote:microphone}}. 
(c) As MFCC is mostly followed by a supervised scheme, the approach may require the voice samples from each user for the correct identification of pitch and tone. However, most of the members in the instantaneous groups are new and appear for the first time. Hence, pre-training is impossible in most of the scenarios.

In this paper, we apply \textit{Complex Cepstrum } (CCEP) to perform tone extraction. CCEP of a signal $\mathcal{S}$ is computed as follows.
\begin{equation}
\text{CCEP}(\mathcal{S}) = \text{IFT}(\log(\text{FT}(\mathcal{S})) + j2{\pi}\ell)
\end{equation}
where FT(.) is the Fourier transform, IFT(.) is the inverse Fourier transform and $j = \sqrt{-1}$. The imaginary part uses complex logarithmic function, and $\ell$ is an integer which is required to properly unwrap the imaginary part of the complex log function. CCEP uses the log compression of the power spectrum, and therefore is less affected by the environmental noise, the nonlinearity of amplifiers and the effect of AGC circuits. To extract the tone from an audio signal, we segment the signal into one second units, and then compute the CCEP for the audio segments. The CCEP for segment $\bar{t}$ from subject $u_i$ is denoted as $cep_i^{\bar{t}}$. These CCEP values for all the subjects are then used for tone similarity measure, as discussed next.

\subsection{Computing Acoustic Context Feature $C^t_{ij}$}
We compute cross-correlation between the CCEP values to measure tone similarity, thereby high and low cross-correlation indicates similar and dissimilar acoustic context between the pair of subjects, respectively. Let $cep_i^{\bar{t}}$ and $cep_j^{\bar{t}}$ denote the CCEP for segment $\bar{t}$ from two different subjects $u_i$ and $u_j$. We compute the segment wise cross-correlation between $cep_i^{\bar{t}}$ and $cep_j^{\bar{t}}$ as $cor_{ij}^{\bar{t}}$, and then average it over the time span $t$. This audio cepstrum cross-correlation is used as the acoustic context similarity $\mathcal{C}_{ij}^t$ for subject pair $u_i$ and $u_j$ during time duration $t$.

\begin{figure}
	\centering
	\captionsetup{justification=centering}
	\includegraphics[clip,height=80pt,keepaspectratio]{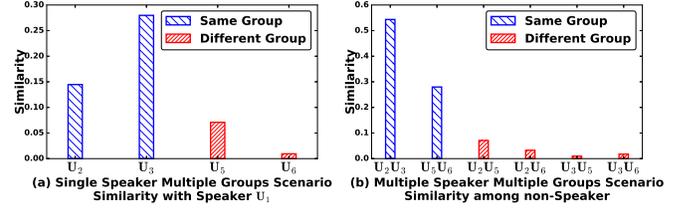}
	\caption{Audio Similarity Variation in Different Scenarios}
	\label{fig:audioScenarioWise}
\end{figure}

To demonstrate the role of tone similarity to compute acoustic context of meeting groups, we consider the groups $G_6$ and $G_7$ formed in outdoor roadside where the group scenario shown in Figure \ref{fig:audioGroup}. Subjects $U_1$, $U_2$, $U_3$ and $U_4$, $U_5$, $U_6$ form the Group $G_6$ and $G_7$, respectively. In the first scenario, subject $U_1$ in group $G_6$ is the dominating speaker, whereas members of Group $G_7$ are mostly silent. Figure \ref{fig:audioScenarioWise}a shows the pairwise context similarity between the individual subject and the dominating speaker $U_1$ (of group $G_6$). We observe that subjects in group $G_6$ (say, $U_2$ and $U_3$) exhibit higher similarity with dominating speaker $U_1$ as compared with the members of the group $G_7$ (say, subjects in $U_5$ and $U_6$). Next, we consider two dominating speakers $U_1$ and $U_4$ in two respective groups $G_6$ and $G_7$. We compute the context similarity between any pair of (non-speaking) subjects. In Figure \ref{fig:audioScenarioWise}b we observe that members belonging to the same group (say $U_2$ and $U_3$ in Group $G_6$ and $U_5$ and $U_6$ in Group $G_7$) exhibit higher context similarity compared to non-group pairs. Precisely, the context similarity between the intragroup members is substantially higher (close to $1.0$) than the intergroup members (close to $0.0$). This result indicates that acoustic context within a single group exhibits substantial similarity.

\begin{figure}
	\captionsetup{justification=centering}
	\begin{minipage}{0.5\linewidth}
		\centering
		\includegraphics[clip,height=100pt,keepaspectratio]{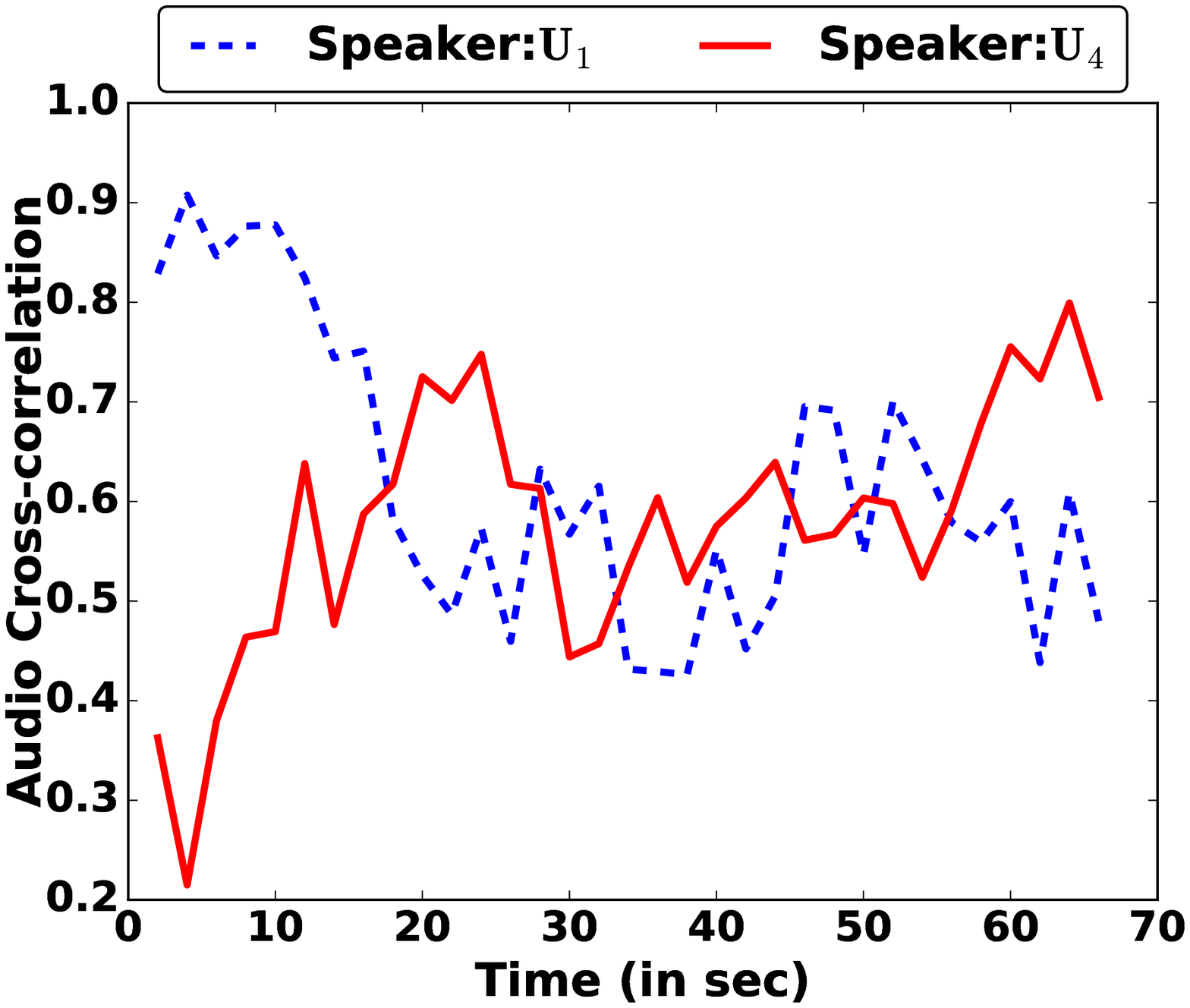}
		\caption{Audio Cross-\\correlation variation over\\time with the moving $U_2$}
		\label{fig:strength}
	\end{minipage}
	\begin{minipage}{0.5\linewidth}
		\centering
		\includegraphics[clip,height=110pt,keepaspectratio]{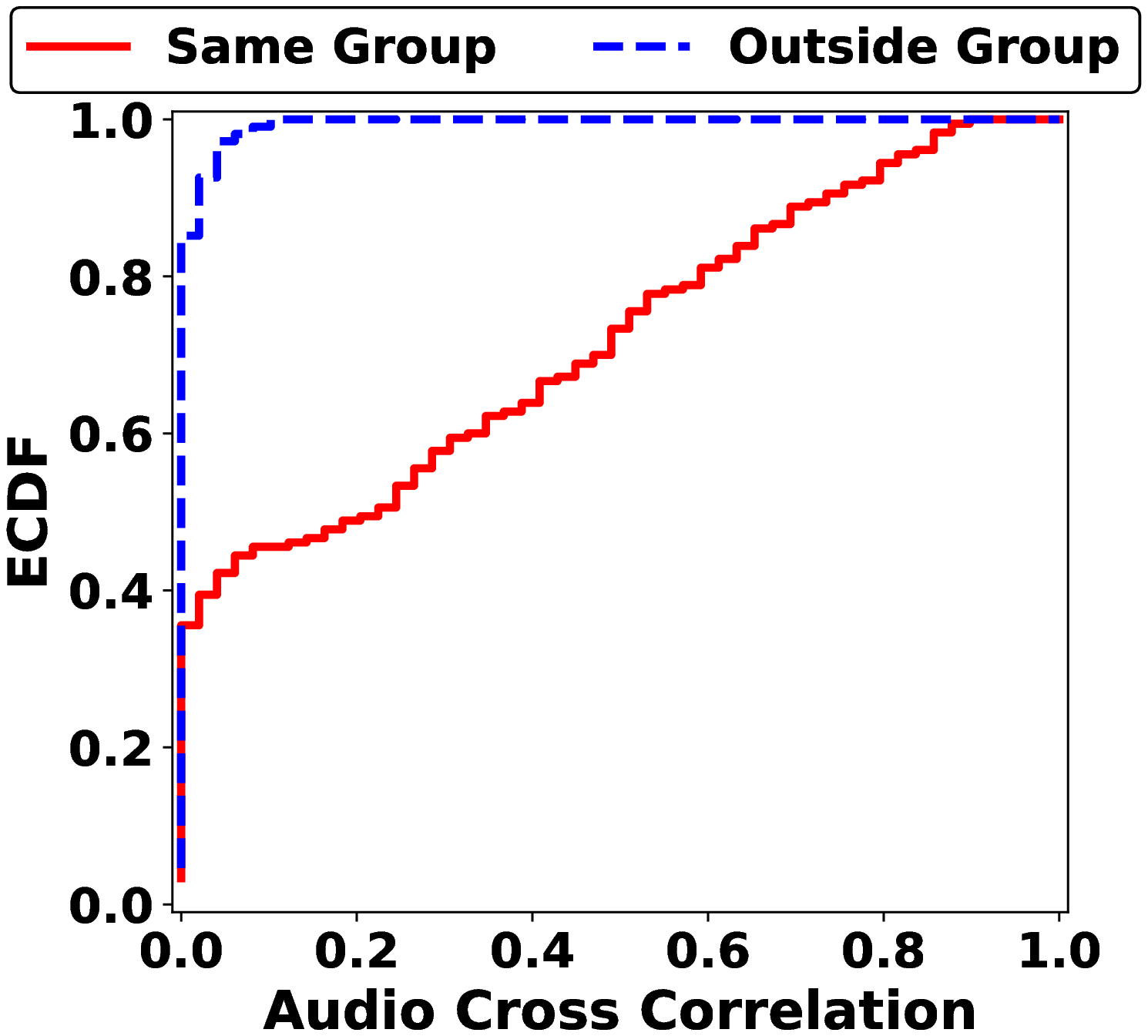}
		\caption{ECDF of Cepstrum Cross-Correlation}
		\label{fig:audioECDF}
	\end{minipage}
\end{figure}

We also investigate the impact of location of a subject on her acoustic context. We set up two groups $G_6$ and $G_7$, $18$m apart in the outdoor environment, with two dominating speakers namely $U_1$ and $U_4$ respectively. We consider one moving subject $U_2$, initially inside the $G_6$ (from Table \ref{table:pilot}) and walks towards group $G_7$ (it takes around $66$ sec to reach group $G_7$ from $G_6$). Figure \ref{fig:strength} shows the variation in the acoustic context similarity between the subject $U_2$ and the dominating speaker over time. We observe that, when the subject is in group $G_6$, the context similarity between the $U_1$ and the subject is high as compared with the $U_4$. The reverse behaviour is noticed at the end of the experiment when the subject reaches $G_7$. However, the context is confusing as the subject located in the middle of both the groups.

%The actual group participation of the subject is determined using the \textit{MeetSense} model (discussed in Section \S\ref{gdm}).

Finally, we perform an overall evaluation, considering all the meeting groups formed in the pilot study. In Figure \ref{fig:audioECDF}, we plot the empirical cumulative distribution (ECDF) of the acoustic context similarity for the pair of subjects. We observe that when the subject pair is in the same group, the acoustic context similarity is high. On the contrary, when one subject is outside the group, context similarity exhibits low value. This establishes the fact that tone similarity, computed from cepstrum cross-correlation, reflects the acoustic context of a group and more importantly, the acoustic context within a single group exhibits substantial similarity.

The aforesaid methodology of extracting acoustic context from smartphone microphone has three broad advantages. First, as the feature is extracted from the dominating tone in an audio signal (captured by cepstrum), it is sufficient if at least one subject in a group talks for a duration. The method can be utilized to consider subjects who belong to a group, but do not prefer to interact (consider a conference presentation); as long as some other subject from that group speaks. Second, the methodology does not violate the privacy of individual subjects. We extract only the tone information from the captured audio signal and do not leverage the exact conversation. Third, the proposed model is unsupervised. Therefore, there is no need of pre-training of the tone information of the group members.

\section{Design of \emph{G\lowercase{roup}S\lowercase{ense}}}\label{gdm}
\begin{figure*}
	\centering
	\includegraphics[clip,width=1.0\textwidth,keepaspectratio]{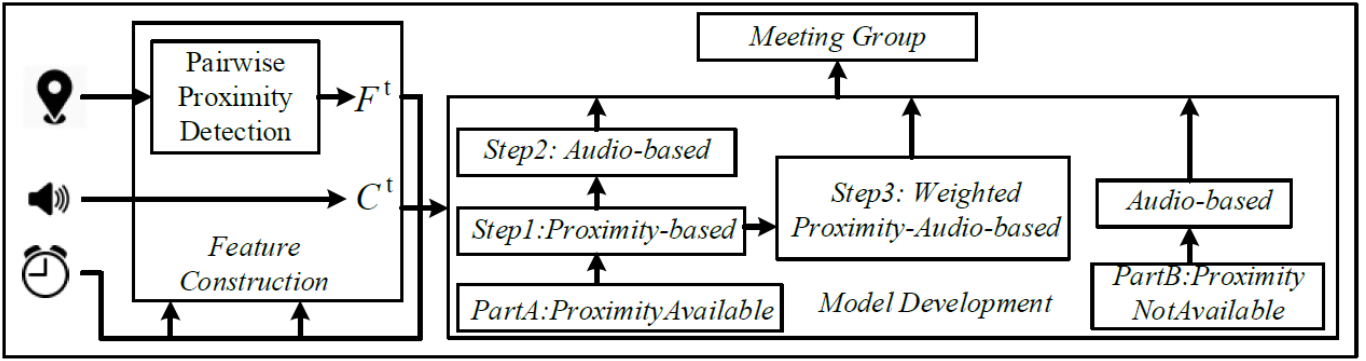}
	\caption{\textit{MeetSense} Model [$\mathcal{F}^t$: Proximity Feature, $\mathcal{C}^t$: Acoustic Context Feature]}
	\label{fig:group_model}
\end{figure*}

\emph{MeetSense} is an unsupervised framework for detecting meeting groups based on subject proximity and acoustic context. Figure~\ref{fig:group_model} shows the flow outline of the \emph{MeetSense} framework. First of all, the sensor logger module records the microphone data along with the proximity indicators followed by the pairwise feature computation. Leveraging on the proximity and acoustic context features, we develop \emph{MeetSense} for meeting group detection.

%\subsection{Sensor Recording and Preprocessing}
%In practice, the uncertainty in the data collection procedure results in routinely dropped readings. We implement a `smoothing filter' to correct the dropped readings and provide clean data. The sliding window $\Delta t$ moves over the data points and interpolates for lost readings collected from the WiFi AP within the time window $\Delta t$. The sliding window size is adaptively fixed based on the observed readings \cite{jeffery2006adaptive}. On the other side, interpolation in microphone data distorts the signal \cite{marvasti1989iterative}. Hence, we extricate the microphone data from window based preprocessing and use the actual signal for feature computation.

\subsection{Feature Construction}
In this module, we compute acoustic context similarity $\mathcal{C}^t_{ij}$ between the subject pair $u_i$ and $u_j$ at time $t$ from the collected microphone log; the detailed computation procedure of acoustic context feature has already been explained in section \ref{fim}. The pairwise proximity similarity feature $\mathcal{F}^t_{ij}$ at time $t$ can be extracted from any of the state of the art techniques \cite{kar63969,sapiezynski2017}. Now considering a subject pair $u_i$ and $u_j$, we need to compute the aggregated features $\overline{\mathcal{F}_{ij}}$ and $\overline{\mathcal{C}_{ij}}$ respectively for time duration $T$. One simplest way of aggregation is computing the mean $\overline{\mathcal{F}_{ij}}$ and $\overline{\mathcal{C}_{ij}}$ from the instantaneous features $\mathcal{F}^t_{ij}$ and $\mathcal{C}^t_{ij}$ respectively for time duration $T$.
However, the signal sample collected from proximity indicator and microphone may suffer from sensitivity and fluctuations.
%For instance, multiple APs, with entirely different transceiving characteristics, may interfere and distort the WiFi signal.
Additionally, the audio signals can also get muffled by obstacles, clothing materials, and are also impacted by the interference. Evidently, the colluded mean features, computed from all the feature points $\mathcal{F}^t_{ij}$ and $\mathcal{C}^t_{ij}$ for the time duration $T$, may not provide a clear indication of feature similarities between the subject pair $u_i$ and $u_j$.
Hence, we compute the refined mean features $\overline{\mathcal{F}_{ij}}$ and $\overline{\mathcal{C}_{ij}}$, by eliminating the low-frequency noise component.
Here we split all the features points (say, for proximity feature $\mathcal{F}^t_{ij}$) into two clusters (via k-means clustering, with $p_{value}<0.05$). Eliminating the minor cluster as the noisy component, we compute the mean $\overline{\mathcal{F}_{ij}}$ from the feature points in the major cluster (see  Algorithm \ref{Algo:feature_group}). However, in case of $p_{value}\geq 0.05$, we compute the mean $\overline{\mathcal{F}_{ij}}$ considering all the feature points in the single cluster.
Similarly, we compute the refined mean acoustic context feature $\overline{\mathcal{C}_{ij}}$ from Algorithm \ref{Algo:feature_group}.

\begin{algorithm}
	\scriptsize
	\caption{Feature\_Construction}
	\raggedright \textbf{Inputs: $\mathbb{F}_{ij}, T$} \\
	\textbf{Output: $\overline{\mathcal{F}_{ij}}$}
	\begin{algorithmic}[1]
		\State $[Cl_1,Cl_2] \leftarrow kmeans((\mathbb{F}_{ij}, T),2)$
		\If {$p_{value} > 0.05$} \Comment{\textit{Single Cluster Scenario}}
		\State $\overline{\mathcal{F}_{ij}} \leftarrow 1/|\mathbb{F}_{ij}|\sum_{\forall \mathcal{F}_{ij}^t} \mathcal{F}_{ij}^t$
		\Else
		\If {$|Cl_1| > |Cl_2|$} \Comment{\textit{Major Cluster} $Cl_1$ \textit{Scenario}}
		\State $\overline{\mathcal{F}_{ij}} \leftarrow 1/|\mathbb{F}_{ij}|\sum_{\forall \mathcal{F}_{ij}^t \in Cl_1} \mathcal{F}_{ij}^t$
		\Else \Comment{\textit{Major Cluster} $Cl_2$ \textit{Scenario}}
		\State $\overline{\mathcal{F}_{ij}} \leftarrow1/|\mathbb{F}_{ij}|\sum_{\forall \mathcal{F}_{ij}^t \in Cl_2} \mathcal{F}_{ij}^t$
		\EndIf
		\EndIf             
	\end{algorithmic}
	\label{Algo:feature_group}
\end{algorithm}

\subsection{Model Development}
Finally, leveraging on the aforementioned features, we develop an unsupervised model for meeting group detection. The model executes Part \emph{A} (Algorithm \ref{Algo:ovaplapAP}) or Part \emph{B} (Algorithm \ref{Algo:noOverlapAP}) depending on the availability of the location information (WiFi, Bluetooth, GPS etc). If the subject possesses location information, the model exploits both proximity as well as acoustic features in Part \textit{A}. Otherwise, the model only relies on the acoustic information in Part \textit{B}. The outcome of the model is all the groups detected by both the individual parts. The outline of the model is described in Algorithm \ref{Algo:group_detection}.

\begin{algorithm}
	\scriptsize
	\caption{MeetSense: Group\_Dection\_Algorithm}
	\raggedright \textbf{Inputs: $u_i(p_i^t,\alpha_i^t) \forall u_i \in \mathbb{U}, \delta_{p_1}, \delta_{p_2}, \delta_\alpha$} \\
	\textbf{Output: $\mathbb{G}^T$}
	\begin{algorithmic}[1]
		\If {${p}_{i}^t \ne \emptyset$}
		\State $\mathbb{U}_o \leftarrow \mathbb{U}_o \cup u_i$
		\EndIf
		\State $\mathbb{U}_{no} \leftarrow \mathbb{U} - \mathbb{U}_o$
		\If {$\mathbb{U}_o \ne \emptyset$} \Comment{\textit{Proximity Available Scenario}}
		\State $\mathbb{G}_o^T \leftarrow$ ProximityAvailable\_Function ($u_i(p_i^t,\alpha_i^t) \forall u_i \in \mathbb{U}_o, \delta_{p_1}, \delta_{p_2}, \delta_\alpha$)
		\EndIf
		\If {$\mathbb{U}_{no} \ne \emptyset$} \Comment{\textit{Proximity Not Available Scenario}}
		\State $\mathbb{G}_{no}^T \leftarrow$ ProximityNotAvailable\_Function ($u_i(\alpha_i^t) \forall u_i \in \mathbb{U}_{no}, \delta_\alpha$)
		\EndIf
		\State $\mathbb{G}^T \leftarrow \mathbb{G}_o^T \cup \mathbb{G}_{no}^T$
	\end{algorithmic}
	\label{Algo:group_detection}
\end{algorithm}

%\begin{algorithm}
%    \caption{MeetSense: Group\_Dection\_Algorithm}
%    \raggedright \textbf{Inputs: $u_i(p_i^t,\alpha_i^t) \forall u_i \in \mathcal{U}, \delta_{p_1}, \delta_{p_2}, \delta_\alpha$} \\
%    \textbf{Output: $\mathbb{G}^T$}
%    \begin{algorithmic}[1]
%%        \State Smooth $p_i^t \leftrightarrow p_i^{t+\Delta t} \forall u_i, t$ \Comment{\textit{Data Smooth}}
%        \State $\mathbb{IS}_{ij}^t \leftarrow p_i^t \cap p_j^t$ $\forall u_i, u_j, t$
%        \If {$\mathbb{IS}_{ij}^t \ne \emptyset$}
%        \State $\mathbb{U}_o \leftarrow \mathbb{U}_o \cup u_i \cup u_j$
%        \EndIf
%        \State $\mathbb{U}_{no} \leftarrow \mathbb{U} - \mathbb{U}_o$
%        \If {$\mathbb{U}_o \ne \emptyset$} \Comment{\textit{Overlapping AP Scenario}}
%        \State $\mathbb{G}_o^T \leftarrow$ OverlapAP\_Function($u_i(p_i^t,\alpha_i^t) \forall u_i \in \mathbb{U}_o, \delta_{p_1}, \delta_{p_2}, \delta_\alpha$)
%        \EndIf
%        \If {$\mathbb{U}_{no} \ne \emptyset$} \Comment{\textit{No Overlapping AP Scenario}}
%        \State $\mathbb{G}_{no}^T \leftarrow$ NoOverlapAP\_Function($u_i(p_i^t,\alpha_i^t) \forall u_i \in \mathbb{U}_{no}, \delta_\alpha$)
%        \EndIf
%        \State $\mathbb{G}^T \leftarrow \mathbb{G}_o^T \cup \mathbb{G}_{no}^T$
%    \end{algorithmic}
%    \label{Algo:group_detection}
%\end{algorithm}

\textbf{Part A:}
%[BM: note the difference between cluster and group......cluster is a potential (but not necessarily a) group]
In this part, we first attempt to extract the cluster of co-locating subjects only based on the pair-wise proximity similarity. If we identify a \emph{highly cohesive cluster} $\mathbb{G}_p^T$ based on proximity only, we consider $\mathbb{G}_p^T$ as a potential meeting group and execute the second step. In the second step, we leverage only on the acoustic context features to detect meeting group(s) $\mathbb{G}_\alpha^T$ from the identified proximity clusters $\mathbb{G}_p^T$. However, if we identify \emph{moderately cohesive cluster} $\mathbb{G}_p^T$ from the proximity features, the model abandons the cluster $\mathbb{G}_p^T$, considering proximity as a critical albeit weak signal, and moves to the third step. In the third step, we combine both the proximity and acoustic context similarity features together to detect cohesive cluster $\mathbb{G}_\alpha^T$ on the complete proximity available population $\mathbb{U}_o$. If $\mathbb{G}_\alpha^T$ exhibits high cohesivity, we assert the cluster $\mathbb{G}_\alpha^T$ as the meeting group. Poor cohesivity in any step rejects the existence of any group in the population. The overall procedure is illustrated in Algorithm \ref{Algo:ovaplapAP}. In the following, we introduce cohesive cluster detection, which is the core of the model.

\textbf{Detection of cohesive cluster:}
Consider a weighted network $\mathcal{CG}(\mathbb{U}, \mathbb{E})$, where $u_i\in\mathbb{U}$ is a subject and $\{e_{ij}, w^e_{ij}\}\in \mathbb{E}$ denotes the weighted link $e_{ij}$ between the subject pair $u_i$ and $u_j$. We apply community detection algorithm \cite{lancichinetti2008benchmark} on $\mathcal{CG}$ to obtain a partition $\mathcal{K}=\{K_1, K_2,\dots,K_m\}$ on population $\mathbb{U}$. 
Essentially the community detection algorithm partitions the network into communities ensuring dense connections within a community and sparser connections between communities. We consider the detected community $K_i$ as a cluster of population $\mathbb{U}$. The cohesivity of the partition $\mathbb{K}$ can be measured with modularity index $\mathcal{M}$ as $\mathcal{M} = \frac{1}{4\varphi} \sum_{ij} (w^e_{ij} - \frac{\rho_i\rho_j}{2\varphi})f(\sigma_i,\sigma_j)$, which reflects the fraction of the links that fall within a given community, compared to the expected fraction if links are distributed at random \cite{lancichinetti2008benchmark}. The $\varphi$, $\rho_i$, $\sigma_i$, and $f(.)$ represent the sum of all of the edge weights in the network, sum of the edge weight attached to node $u_i$, the community of node $u_i$, and delta function, respectively. 
Notably, modularity of a weighed fully connected graph becomes $zero$ if all the nodes form a single large community \cite{newman2006modularity}. In this paper, we apply Walktrap algorithm \cite{pons2005computing}; however, our methodology is not sensitive to any specific (weighted) community detection algorithm.

Algorithm \ref{Algo:ovaplapAP} comprises of the following three steps.

\textbf{Step 1:}
We construct a complete proximity graph $\mathcal{PG}(\mathbb{U}_o, \overline{\mathbb{F}})$ where $\mathbb{U}_o$ denotes the complete proximity available population and $\{e_{ij}, \overline{\mathcal{F}_{ij}}\}\in\overline{\mathbb{F}}$ is a link between the subject pair $u_i$ and $u_j$ weighted by the proximity feature $\overline{\mathcal{F}_{ij}}$ computed over the time $T$. We apply community detection algorithm on the proximity graph $\mathcal{PG}$ to discover the cluster $\mathbb{G}^T_p$ with modularity $\mathcal{M}_p$.
If the $\mathcal{M}_p$ is above a threshold $\delta_{p_1}$, we consider $\mathbb{G}^T_p$ as the candidate meeting group and move to step 2. If $\mathcal{M}_{p}$ falls below a threshold $\delta_{p_2}$, we reject the existence of any meeting group in population $\mathbb{U}$. Otherwise, we move to step 3.

\textbf{Step 2:}
We construct the complete acoustic context graphs $\mathcal{IG}(\mathbb{G}_p^T, \overline{\mathbb{C}})$ where $\{e_{ij}, \overline{\mathcal{C}_{ij}}\}\in\overline{\mathbb{C}}$ links between subject pair $u_i$ and $u_j \in \mathbb{G}_p^T$. Essentially, in $\mathcal{IG}$, the link weight $\overline{\mathcal{C}_{ij}}$ depicts the acoustic context similarity over the time $T$. Similar to step 1, we apply the community detection on $\mathcal{IG}$ to discover the cluster $\mathbb{G}^T_\alpha$ with modularity $\mathcal{M}_\alpha$. If the $\mathcal{M}_\alpha$ is above a threshold $\delta_{\alpha}$, we confirm $\mathbb{G}^T_\alpha$ as the detected meeting groups. Otherwise, we reject the existence of meeting groups in population $\mathbb{U}_o$.

\textbf{Step 3:}
We construct a complete proximity-acoustic context graph $\mathcal{MG}(\mathbb{U}_o, \mathbb{W})$ where $\{e_{ij}, \mathcal{W}_{ij}\} \in \mathbb{W}$ links between subject pair $u_i$ and $u_j \in \mathbb{U}_o$ weighted by $\mathcal{W}_{ij}=(1-w) \times \overline{\mathcal{F}_{ij}}+w \times \overline{\mathcal{C}_{ij}}$. Essentially, in $\mathcal{MG}$, the link weight $\mathcal{W}_{ij}$ carries the information from both acoustic context and proximity feature. Similar to step 1, we apply the community detection on $\mathcal{MG}$ to discover the cluster $\mathbb{G}^T_w$ with modularity $\mathcal{M}_w$. If the $\mathcal{M}_w$ is above a threshold $\delta_{\alpha}$, we confirm $\mathbb{G}^T_w$ as the detected meeting groups. Otherwise, we reject the presence of any group in population $\mathbb{U}_o$.

\begin{algorithm}
	\scriptsize
	\caption{ProximityAvailable\_Function}
	\raggedright \textbf{Inputs: $u_i(p_i^t,\alpha_i^t) \forall u_i \in \mathbb{U}_o, \delta_{p_1}, \delta_{p_2}, \delta_\alpha$} \\
	\textbf{Output: $\mathbb{G}^T$}
	\begin{algorithmic}[1]
		\State Compute $\mathcal{F}_{ij}^t, \mathcal{C}_{ij}^t$ \Comment{\textit{Feature Generation}}
		\State $\overline{\mathcal{F}_{ij}} \leftarrow$ Feature\_Construction  ($\mathbb{F}_{ij}, T$) $\forall u_i, u_j$
		\State $(\mathbb{G}_p^T, \mathcal{M}_p) \leftarrow $ Community\_Detection ($\mathbb{U}_o,\overline{\mathbb{F}}$)
		\If {$\mathcal{M}_p \ge \delta_{p_1}$} \Comment{\textit{Proximity Dominating Scenario}}
		\State $\overline{\mathcal{C}_{ij}} \leftarrow$ Feature\_Construction  ($\mathbb{C}_{ij}, T$) $\forall u_i, u_j \in \mathcal{G}, \mathcal{G} \in \mathbb{G}_p^T$     
		\State $(\mathbb{G}_\alpha^T, \mathbb{M}_\alpha) \leftarrow $     Community\_Detection ($\mathbb{G}_p^T,\overline{\mathbb{C}}$)
		\If {$\mathcal{M}_\alpha \ge \delta_\alpha \forall (\mathcal{G}_\alpha,\mathcal{M_\alpha}) \in (\mathbb{G}_\alpha^T,\mathbb{M}_\alpha)$} \Comment{\textit{Proximity \& Audio Influence Scenario}}
		\State $\mathbb{G}^T \leftarrow \mathbb{G}^T \cup \mathcal{G}_\alpha$
		\Else \Comment{\textit{Proximity Influence \& Audio Insignificance Scenario}}
		\State Failure
		\EndIf
		\Else
		\If {$\mathcal{M}_p < \delta_{p_1} and \mathcal{M}_p \ge \delta_{p_2}$}
		\State $\overline{\mathcal{C}_{ij}} \leftarrow$ Feature\_Construction  ($\mathbb{C}_{ij}, T$) $\forall u_i, u_j$
		\State $(\mathbb{G}_w^T, \mathcal{M}_w) \leftarrow $ Community\_Detection ($\mathbb{U}_o,(1 - w) \times \overline{\mathbb{F}} + w \times \overline{\mathbb{C}}$) $\forall w \in [0,1]$ \Comment{\textit{Weighted Features}}
		\State $(\mathbb{G}_w^T, \mathcal{M}) \leftarrow max(\mathbb{G}_w^T, \mathcal{M}_w) \forall w \in [0,1]$
		\If {$\mathcal{M} \ge \delta_\alpha$} \Comment{\textit{Proximity Confused \& Audio Influence Scenario}}
		\State $\mathbb{G}^T \leftarrow  \mathbb{G}_w^T$
		\Else \Comment{\textit{Proximity Confused \& Audio Insignificance Scenario}}
		\State Failure
		\EndIf
		\Else \Comment{\textit{Proximity Insignificance Scenario}}
		\State Failure
		\EndIf
		\EndIf
	\end{algorithmic}
	\label{Algo:ovaplapAP}
\end{algorithm}

\textbf{Part B:}
Due to the unavailability of location data, in this part, we completely rely on the acoustic context similarity between the subjects. We first construct a complete acoustic context graph $\mathcal{IG}(\mathbb{U}_{no}, \overline{\mathbb{C}})$ where $\{e_{ij}, \overline{\mathcal{C}_{ij}}\} \in \overline{\mathbb{C}}$ links between subject pair $u_i$ and $u_j \in \mathbb{U}_{no}$. Essentially, in $\mathcal{IG}$, the link weight $\overline{\mathcal{C}_{ij}}$ carries the information of the acoustic context feature over the time $T$. Similar to part \textit{A}, we apply the community detection on $\mathcal{IG}$ to discover the clusters $\mathbb{G}^T_\alpha$ with modularity $\mathcal{M}_\alpha$. If the $\mathcal{M}_\alpha$ is above a threshold $\delta_{\alpha}$, we confirm $\mathbb{G}^T_\alpha$ as the detected meeting groups. Otherwise, we reject the existence of meeting groups in population $\mathbb{U}_{no}$. The outline of the overall mechanism is portrayed in Algorithm \ref{Algo:noOverlapAP}.

\begin{algorithm}
	\scriptsize
	\caption{ProximityNotAvailable\_Function}
	\raggedright \textbf{Inputs: $u_i(\alpha_i^t) \forall u_i \in \mathbb{U}_{no}, \delta_\alpha$} \\
	\textbf{Output: $\mathbb{G}^T$}
	\begin{algorithmic}[1]
		\State Compute $\mathcal{C}_{ij}^t$ \Comment{\textit{Feature Generation}}
		\State $\overline{\mathcal{C}_{ij}} \leftarrow$ Feature\_Construction ($\mathbb{C}_{ij}, T$) $\forall u_i, u_j$
		\State $(\mathbb{G}_\alpha^T, \mathcal{M}_\alpha) \leftarrow $ Community\_Detection($\mathbb{U}_{no},\overline{\mathbb{C}}$)
		\If {$\mathcal{M}_\alpha \ge \delta_\alpha$} \Comment{\textit{Audio Influence Scenario}}
		\State $\mathbb{G}^T \leftarrow  \mathbb{G}_\alpha^T$
		\Else \Comment{\textit{Audio Insignificance Scenario}}
		\State Failure
		\EndIf
	\end{algorithmic}
	\label{Algo:noOverlapAP}
\end{algorithm}

%\begin{algorithm}
%    \caption{NoOverlapAP\_Function}
%    \raggedright \textbf{Inputs: $u_i(p_i^t,\alpha_i^t) \forall u_i \in \mathbb{U}_{no}, \delta_\alpha$} \\
%    \textbf{Output: $\mathbb{G}^T$}
%    \begin{algorithmic}[1]
%        \If {$|p_i^t| = 0$ or $|p_j^t| = 0$} \Comment{\textit{No WiFi Coverage Scenario}}
%            \State Compute $\mathcal{C}_{ij}^t$ \Comment{\textit{Feature Generation}}
%            \State $\overline{\mathcal{C}_{ij}} \leftarrow$ Feature\_Grouping ($\mathbb{C}_{ij}, T$) $\forall u_i, u_j$
%            \State $(\mathbb{G}_\alpha^T, \mathcal{M}_\alpha) \leftarrow $ Community\_Detection($\mathbb{U}_{no},\overline{\mathbb{C}}$)
%            \If {$\mathcal{M}_\alpha \ge \delta_\alpha$} \Comment{\textit{Audio Influence Scenario}}
%                \State $\mathbb{G}^T \leftarrow  \mathbb{G}_\alpha^T$
%            \Else \Comment{\textit{Audio Insignificance Scenario}}
%                \State Failure
%            \EndIf
%        \Else \Comment{\textit{No Common AP Scenario}}
%            \State No Group
%        \EndIf
%    \end{algorithmic}
%    \label{Algo:noOverlapAP}
%\end{algorithm}

\section{Performance Evaluation} \label{ds}
We evaluate \textit{MeetSense} by developing a smartphone-based application and deploying it over IIT Kharagpur campus spreading $8.5$ square kilometres, consisting of administrative blocks, approximately $30$ academic departments along with campus residential, hostels and market areas. We first discuss the implementation of \textit{MeetSense} followed by the field study and performance comparison with different baselines. 

\subsection{Field Study and Data Collection}
The data collection for \textit{MeetSense} is done initially through an Android app, \emph{DataGatherer}, which has been launched over the smartphones of $40$ subjects consisting of undergraduate and postgraduate students, summer interns, research scholars and faculties of the institute. In our implementation of \textit{MeetSense}, we have considered $T \geq 15 \text{mins}$, that means if an interaction continues for at least $15$ minutes, we consider it as a group. Nevertheless, this is an application specific tunable parameter. We have used different models of smartphones, where costs per phone range from USD $150\$$ to USD $700\$$ approximately. We primarily gather WiFi (BSSID and signal strength) and audio data from smartphones through \emph{DataGatherer} which sends the data to a central server. The WiFi data is used for proximity detection based on existing methodologies~\cite{sapiezynski2017,kar63969}, and then audio data is used to detect the groups among the subjects in proximity. The app scans the available WiFi access points once in a minute time interval, and continuous audio signals are tracked at a sampling rate of $44.1$ kHz for a minute time span followed by an interval of three minutes. Moreover, we have discarded the details of the access points having the signal strength less than $-80$dBm which is the minimum signal strength for basic connectivity\footnote{https://support.metageek.com/hc/en-us/articles/201955754-Understanding-WiFi-Signal-Strength (Accessed on Apr 12, 2018)}. The data has been collected for approximately six months. 

We collect the ground truth meeting group information from the participants for validation. In ground truth data collection, a questionnaire app periodically probes from the participants regarding the (a) start time of the meeting, (b) end time of the meeting, (c) meeting venue and (d) details of the other participants of the meeting. In some cases where a participant misses to provide the ground truth information, we validate the detected meeting groups from the participants by forwarding an email at every two hours of each day. Based on the field study collected data, we identify seven typical meeting group scenarios, which repeatedly occurred (at least once in a week) during the six months field study. These situations are highlighted keeping in mind the critical conditions of group formation that were developed in the pilot study (Section~\ref{so}); thus reflect realistic meeting group scenarios with high probability. We evaluate the performance of \textit{MeetSense} and compare it with other baselines considering these typical scenarios, as well as different scenarios observed from the collected data. 
%The different meeting group scenarios are shown in Figure \ref{fig:group}. 
These scenarios are as follows.

\noindent\textbf{S1 (Indoor: Two groups at neighbouring rooms)}: $3$ subjects attend a lecture in classroom C-$119$, and $2$ subjects have another meeting in the FV Lab opposite to C-$119$ at the same instance of time.

\noindent\textbf{S2 (Indoor: Three groups at different rooms at the same department)}: $4$ subjects interact in the faculty office in the second floor, $2$ subjects are in a meeting at the departmental library opposite to that faculty office, and $2$ subjects are in another meeting at the SMR Lab in the first floor.

\noindent\textbf{S3 (Outdoor: Cafeteria interactions)}: Two different groups at the cafeteria, one with $3$ subjects in front of the cafeteria and another one with $3$ subjects at the back of the cafeteria.

\noindent\textbf{S4 (Indoor: Large single group)}: $7$ subjects attend a presentation at the departmental conference room.

\noindent\textbf{S5 (Indoor: Two different groups at a large lab)}: $3$ subjects meet at cubicle K-$1$ and another $3$ subjects meet in the cubicle K-$10$ of the SMR Lab.

\noindent\textbf{S6 (Indoor: Two roaming groups)}: $3$ subjects together and $2$ subjects together roam around the corridor of the department, and moves from one room to another, forming two non-static groups.

\noindent\textbf{S7 (Outdoor: Two roaming groups)}: $5$ subjects together and $2$ subjects together roam within the campus maintaining a certain distance from each other, forming two non-static groups.

\subsection{Preprocessing -- Proximity Computation for Group Detection}
%We have analyzed the performance of \textit{MeetSense} with following three baselines. As the baselines are mainly focusing on proximity pair identification, we acquire the idea of those techniques and formulate the rest using our methodologies (detailed in subsection \ref{sec:exp_proc}). 
As we discussed earlier, group detection first requires to find out the subjects in proximity, and in \textit{MeetSense}, we utilize the existing proximity detection mechanisms that have been well studied in the literature. We focus on the two approaches of proximity detection based on WiFi data, as follows. 

\noindent\textbf{(a) Supervised learning with WiFi-based proximity sensing (\textit{SLWP})~\cite{sapiezynski2017}:} Sapiezynski \emph{et al.} developed a WiFi access point based \emph{supervised} proximity detection mechanism, where Bluetooth data is considered as the ground truth. In this approach, a set of WiFi-based features has been computed, such as overlapping access points, signal strength from different access points etc., and then a support vector machine (SVN) is used to classify whether two subjects are in proximity or not. 
%We employ this proximity detection model and augment it with the audio signal sensing methodology used in \textit{MeetSense} for meeting groups detection.

\noindent\textbf{(b) \textit{Next2Me}~\cite{kar63969}:} This is a smartphone-based system for capturing social interactions within close proximity users.  Next2Me uses WiFi signal information for measuring the pairwise co-located Manhattan distance between the users, and then a threshold over the distance function is used to find out the subjects in proximity. It can be noted that this is an unsupervised approach.

\subsection{Baselines for Audio Based Interaction Detection}

We have analyzed the performance of \textit{MeetSense} with following three baselines, which utilize audio signals for acoustic context detection. We use the proximity followed by audio based acoustic context to detect various meeting groups. 

\noindent\textbf{(a) \textit{Next2Me}~\cite{kar63969}:}  After determining the subjects in proximity, Next2Me utilizes Jaccard similarity over top $n$ audio frequencies to capture the audio fingerprints of various subjects. Finally, they generate social community by applying Louvain community detection algorithm.  

\noindent\textbf{(b) \textit{AudioMatch}~\cite{casagranda2015audio}:} Casagranda \emph{et al.} implemented a smartphone based group detection system based on the joint usage of GPS and audio fingerprints. The GPS information is used for filtering out nearby devices. On top of the GPS based clusters, the audio module is executed for identifying the groups. \textit{AudioMatch} uses short time Fourier transform (STFT) with overlapping Hamming window. Finally, it computes the Hamming distance between the pair of devices for detecting the nearby pairs. 

Next, we discuss the experimental procedure by combing the WiFi-based proximity detection and audio based acoustic context detection together. 
\begin{table*}[]
	\centering
	\caption{Performance Comparison}
	\label{table:overall_perf}
	\begin{tabular}{|c|l|l|l|l|l|l|l|l|l|l|l|l|}
		\hline
		\multirow{3}{*}{ID} & \multicolumn{6}{c|}{SLWP}                                                                                                                                                                                                                                                                                                                                                                                                                                                   & \multicolumn{6}{c|}{Next2Me}                                                                                                                                                                                                                                                                                                                                                                                                                                                \\ \cline{2-13} 
		& \multicolumn{2}{c|}{Next2Me}                                                                                                                            & \multicolumn{2}{c|}{GroupSense}                                                                                                                         & \multicolumn{2}{c|}{AudioMatch}                                                                                                                         & \multicolumn{2}{c|}{Next2Me}                                                                                                                            & \multicolumn{2}{c|}{GroupSense}                                                                                                                         & \multicolumn{2}{c|}{AudioMatch}                                                                                                                         \\ \cline{2-13} 
		& \multicolumn{1}{c|}{\begin{tabular}[c]{@{}c@{}}F\_1\\ Score\end{tabular}} & \multicolumn{1}{c|}{\begin{tabular}[c]{@{}c@{}}Modu-\\ larity\end{tabular}} & \multicolumn{1}{c|}{\begin{tabular}[c]{@{}c@{}}F\_1\\ Score\end{tabular}} & \multicolumn{1}{c|}{\begin{tabular}[c]{@{}c@{}}Modu-\\ larity\end{tabular}} & \multicolumn{1}{c|}{\begin{tabular}[c]{@{}c@{}}F\_1\\ Score\end{tabular}} & \multicolumn{1}{c|}{\begin{tabular}[c]{@{}c@{}}Modu-\\ larity\end{tabular}} & \multicolumn{1}{c|}{\begin{tabular}[c]{@{}c@{}}F\_1\\ Score\end{tabular}} & \multicolumn{1}{c|}{\begin{tabular}[c]{@{}c@{}}Modu-\\ larity\end{tabular}} & \multicolumn{1}{c|}{\begin{tabular}[c]{@{}c@{}}F\_1\\ Score\end{tabular}} & \multicolumn{1}{c|}{\begin{tabular}[c]{@{}c@{}}Modu-\\ larity\end{tabular}} & \multicolumn{1}{c|}{\begin{tabular}[c]{@{}c@{}}F\_1\\ Score\end{tabular}} & \multicolumn{1}{c|}{\begin{tabular}[c]{@{}c@{}}Modu-\\ larity\end{tabular}} \\ \hline
		S1                  & 1.0000                                                                    & 0.1124                                                                      & 1.0000                                                                    & 0.2879                                                                      & 0.7273                                                                    & 0.0000                                                                      & 1.0000                                                                    & 0.0412                                                                      & 1.0000                                                                    & 0.2907                                                                      & 0.7273                                                                    & 0.0000                                                                      \\ \hline
		S2                  & 0.9000                                                                    & 0.2030                                                                      & 1.0000                                                                    & 0.1760                                                                      & 0.6667                                                                    & 0.0000                                                                      & 0.9000                                                                    & 0.2030                                                                      & 1.0000                                                                    & 0.1760                                                                      & 0.6667                                                                    & 0.0000                                                                      \\ \hline
		S3                  & 0.5333                                                                    & 0.1261                                                                      & 1.0000                                                                    & 0.3642                                                                      & 0.7273                                                                    & 0.0000                                                                      & 0.5333                                                                    & 0.1261                                                                      & 1.0000                                                                    & 0.3642                                                                      & 0.7273                                                                    & 0.0000                                                                      \\ \hline
		S4                  & 0.8326                                                                    & 0.0772                                                                      & 1.0000                                                                    & 0.0000                                                                      & 1.0000                                                                    & 0.0000                                                                      & 0.8326                                                                    & 0.0772                                                                      & 1.0000                                                                    & 0.0000                                                                      & 1.0000                                                                    & 0.0000                                                                      \\ \hline
		S5                  & 0.8571                                                                    & 0.0732                                                                      & 1.0000                                                                    & 0.3801                                                                      & 0.7273                                                                    & 0.0000                                                                      & 0.8571                                                                    & 0.0732                                                                      & 1.0000                                                                    & 0.3801                                                                      & 0.7273                                                                    & 0.0000                                                                      \\ \hline
		S6                  & 1.0000                                                                    & 0.0000                                                                      & 1.0000                                                                    & 0.0000                                                                      & 1.0000                                                                    & 0.0000                                                                      & 1.0000                                                                    & 0.0000                                                                      & 1.0000                                                                    & 0.0000                                                                      & 1.0000                                                                    & 0.0000                                                                      \\ \hline
		S7                  & 0.5833                                                                    & 0.1942                                                                      & 0.6500                                                                    & 0.0976                                                                      & 0.8333                                                                    & 0.0000                                                                      & 0.5833                                                                    & 0.1942                                                                      & 0.6500                                                                    & 0.0976                                                                      & 0.8333                                                                    & 0.0000                                                                      \\ \hline
		ALL                 & 0.7971                                                                    & 0.0866                                                                      & \textbf{0.9421}                                                           & \textbf{0.2114}                                                             & 0.8212                                                                    & 0.0000                                                                      & 0.7971                                                                    & 0.0826                                                                      & \textbf{0.9421}                                                           & \textbf{0.2116}                                                             & 0.8212                                                                    & 0.0000                                                                      \\ \hline
	\end{tabular}
\end{table*}

\subsection{Experimental Procedure}\label{sec:exp_proc}
%In our experiments, we mainly focus on the identification of meeting group. 
%%Those groups form either inside or outside environment. Hence, 
%We consider WiFi as the proximity modality for our meeting group detection experiments. As we mention earlier, \textit{SLWP} uses only WiFi fingerprint for detecting the pairwise proximity whereas \textit{AudioMatch} explores GPS and audio information. On the other side, \textit{Next2Me} uses WiFi fingerprint for filtering out the far away users from the set of all users. Hence, any of the approaches are not directly aligned with our overall objectives. Therefore, we use the pairwise information on top of the \textit{MeetSense} model in the following manner:
For comparing the performance of \textit{MeetSense} under the different environment with different baselines, we consider following combinations of proximity (P) and acoustic context (I) detection mechanisms. It can be noted that \textit{MeetSense} primarily focuses on capturing acoustic context, whereas the proximity module is borrowed from existing methodologies. The different combinations of proximity and acoustic context detection mechanisms used in our experiments are as follows.   

\noindent\textbf{I. \textit{SLWP (P) + Next2Me (I)}:} In this arrangement, we extract the pairwise proximity information from \textit{SLWP}, and 
%apply \textit{MeetSense} Feature Construction algorithm \ref{Algo:feature_group} followed by community detection module. 
the outcome is directly fed to the \textit{Next2Me} audio model for group detection.

\noindent\textbf{II. \textit{SLWP (P) + MeetSense (I)}:} This arrangement uses the pairwise proximity information from \textit{SLWP}. Then, \textit{MeetSense} audio centric context detection is applied on top of the proximity outcome.   

\noindent\textbf{III. \textit{SLWP (P) + AudioMatch (I)}:} In this arrangement, we apply \textit{SLWP} for pairwise proximity detection. 
%followed by the \textit{MeetSense} feature modules like the previous two setup.
After that, \textit{AudioMatch} is applied on the outcome of the proximity clusters for detecting the pairwise acoustic context from the audio signals. It can be noted that we have not used GPS for proximity detection as used in AudioMatch, as GPS gives very poor signal in the indoor scenarios. However, the audio module is used as it is, and finally, the community detection algorithm is used for group detection. 
%The pairwise interaction information is finally fed to the \textit{MeetSense} Feature Construction algorithm \ref{Algo:feature_group} followed by community detection module for meeting group detection.  

\noindent\textbf{IV. \textit{Next2Me (P) + Next2Me (I)}:} This arrangement is analogous with the actual \textit{Next2Me} system, where both the WiFi based proximity detection and the audio based acoustic context detection, as done in Next2Me, are used for group detection.

\noindent\textbf{V. \textit{Next2Me (P) + MeetSense (I)}:} In this arrangement, we compute the proximity-based pairwise distance following \textit{Next2Me} proximity module. The pairwise similarity is computed by reversing the pairwise distance value. Then, we apply \textit{MeetSense} Feature Construction algorithm \ref{Algo:feature_group} followed by community detection module on the pairwise similarity value. Finally, \textit{MeetSense} acoustic context module is employed on top of the proximity outcome.  

\noindent\textbf{VI. \textit{Next2Me (P) + AudioMatch (I)}:} This arrangement uses the proximity information from \textit{Next2Me} like the previous setup. After that, \textit{AudioMatch} is applied on the outcome of the proximity clusters. The pairwise acoustic context information is finally fed to the community detection algorithm for meeting group detection.

\subsection{\textit{MeetSense} Performance}
We first evaluate the overall performance of \textit{MeetSense} in terms of $F_1$-Score~\cite{f1score} defined as follows. Let $\varGamma$ and $\varUpsilon$ be the sets of meeting groups in the ground truth data and the ones detected by \textit{MeetSense}, respectively. Then $F_{1_{\kappa \nu}}=\frac{2 \times |\kappa \cap \nu|}{|\kappa|+|\nu|}$ where $\kappa \in \varGamma$ and $\nu \in \varUpsilon$. This parameter captures the accuracy of the detected group $\nu$ in terms of membership overlap with ground truth $\kappa$ for the meeting duration $T$. Now, to obtain the final accuracy of \textit{MeetSense} considering all the detected meeting groups, we compute the average $F_1$-Score as $F_1 = \frac{\sum\limits_{\forall \kappa \in \varGamma; \forall \nu \in \varUpsilon} F_{1_{\kappa \nu}}}{|\varUpsilon|}$.

Table~\ref{table:overall_perf} summarizes the performance of \textit{MeetSense} in terms of $F_1$-Score and modularity ($\mathcal{M}$) for seven representative scenarios as well as for all the observed scenarios combined.  
We experimentally set up the model thresholds ($\delta$) based on the best performance of the overall scenarios. 
The modularity index $\mathcal{M}$ indicates the cohesiveness of the detected groups; hence even a low $F_1$-Score with high modularity contributes more to identify maximum participants in a meeting group. Although \textit{MeetSense} performs marginally worse for certain scenarios, like during outdoor mobile groups due to the high environmental noise, we observe that accuracy is more than $90\%$ for most of the cases. 

\subsubsection{Baseline Comparison}
As mentioned earlier, we have set up six different arrangements proposed as well as existing schemes for investigating the performance at different scenarios. Table~\ref{table:overall_perf} compares the performance of \textit{MeetSense} with \textit{AudioMatch} and \textit{Next2Me} with two different proximity schemes. For all the three scenarios when \textit{SLWP} is used for proximity measure, we observe that \textit{MeetSense} outperforms the other baselines. Although \textit{Next2Me} uses audio based features to capture the social interaction among the subjects (which is similar to the acoustic context for \textit{MeetSense}), it uses Jaccard similarity among top $n$ audio frequencies, which is susceptible to environmental noise. For example, in an outdoor environment, the sound frequencies originated from external entities, such as moving vehicles, can fall within the top $n$ frequency components. As a consequence, we observe that although \textit{Next2Me} manages to perform well in indoor scenarios, it poorly performs in the outdoor environment. On the other side, \textit{AudioMatch} applies hamming distance measure of the logarithmic amplitude of the audio signal for suppressing the noise component. Although this scheme works for artificially generated Gaussian noise, it poorly performs in the presence of real environmental noise. The impact is visible in the outdoor scenarios. 

Next, we have tested the scenarios with three baselines for acoustic context measurement along with \textit{Next2Me} proximity measure. The results of these are apparently similar to the results from the scenarios with \textit{SLWP} proximity measure, except for the proximity dominant scenario $S1$ in \textit{MeetSense}. The similar results also claim that the audio features are more dominant compared to the proximity features for meeting group detection. Moreover, this validates the importance of Algorithm \ref{Algo:noOverlapAP} in the overall meeting group detection mechanism.   

%\notesc{Mention that use has used ANOVA and give the citation of ANOVA. Otherwise, how will people understand what is $p$ value?}

\subsubsection{Robustness of Acoustic Context Measure}
\begin{figure}
	\centering
	\captionsetup{justification=centering}
	\includegraphics[clip,width=\columnwidth,keepaspectratio]{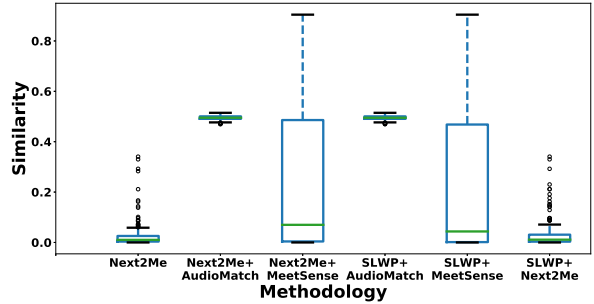}
	\caption{Mean difference of similarity obtained through different methodologies}
	\label{fig:anova}
\end{figure}

For investigating the variations in the performance of different schemes, we report the box-plot of the pairwise feature similarity values for the acoustic context, shown in Figure \ref{fig:anova}. The box plot depicts that there are significant mean differences between the various schemes. In the box plot, the medians for different schemes are shown in red lines. Focusing on the upper and the lower halves from the median, the results show that \textit{MeetSense} captures significant variations in the pairwise similarity between the subjects. As we consider multiple meeting group scenarios, the variation of the pairwise similarity between the subjects is justified. It can be noted that the median is biased towards the lower values because the pairwise feature similarity becomes very close to zero whenever the two subjects in the pair are from different groups. However, a wide variation of similarity values greater than $0.1$ is observed when both the subjects are in the same group. On the contrary, the results for other baselines depict that \textit{Next2Me} and \textit{AudioMatch} show the minimal difference in the upper and the lower halves from the median. Therefore, the constructed feature is incapable of distinguishing between the acoustic context when the subjects are in the same group or different groups. Hence, the $F_1$ score significantly drops for those baselines. Additionally, we also observe that the median value is closer to the first quartile. 
%This signifies that the pairwise similarity value is denser near the zero value. 
As we capture the proximity and audio signatures of the subjects in various environments, the similarity values between each pair of subjects significantly varies over the different meeting groups, causing the dense zone towards the lower halves from the median. The wide variation of the pairwise similarity values in different groups further interprets that the simple thresholding based scheme is not suitable for detecting various types of meeting groups in the diverse environment. Hence, it justifies the requirement of the complex \textit{MeetSense} scheme (Algorithm \ref{Algo:group_detection}).

\subsubsection{Dissecting the Methodologies}
\begin{figure}[t]
	\centering
	\captionsetup{justification=centering}
	\begin{subfigure}[b]{0.24\textwidth}
		\includegraphics[clip,height=120pt,keepaspectratio]{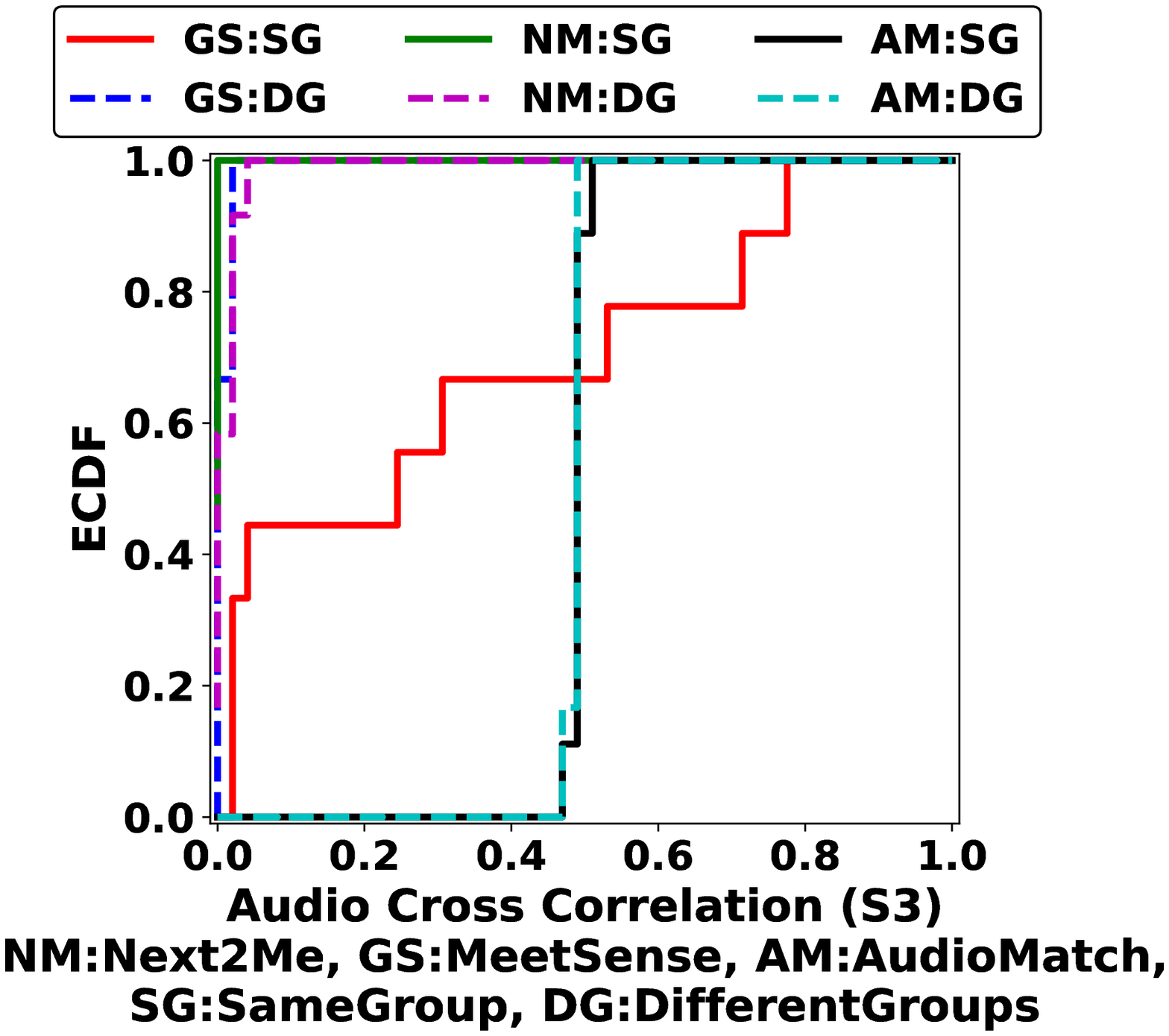}
		\caption{Audio correlation\\comparison for scenario $S3$}
		\label{fig:ecdf_compare_s3}
	\end{subfigure}
	\hfill
	\begin{subfigure}[b]{0.24\textwidth}
		\centering
		\includegraphics[clip,height=120pt,keepaspectratio]{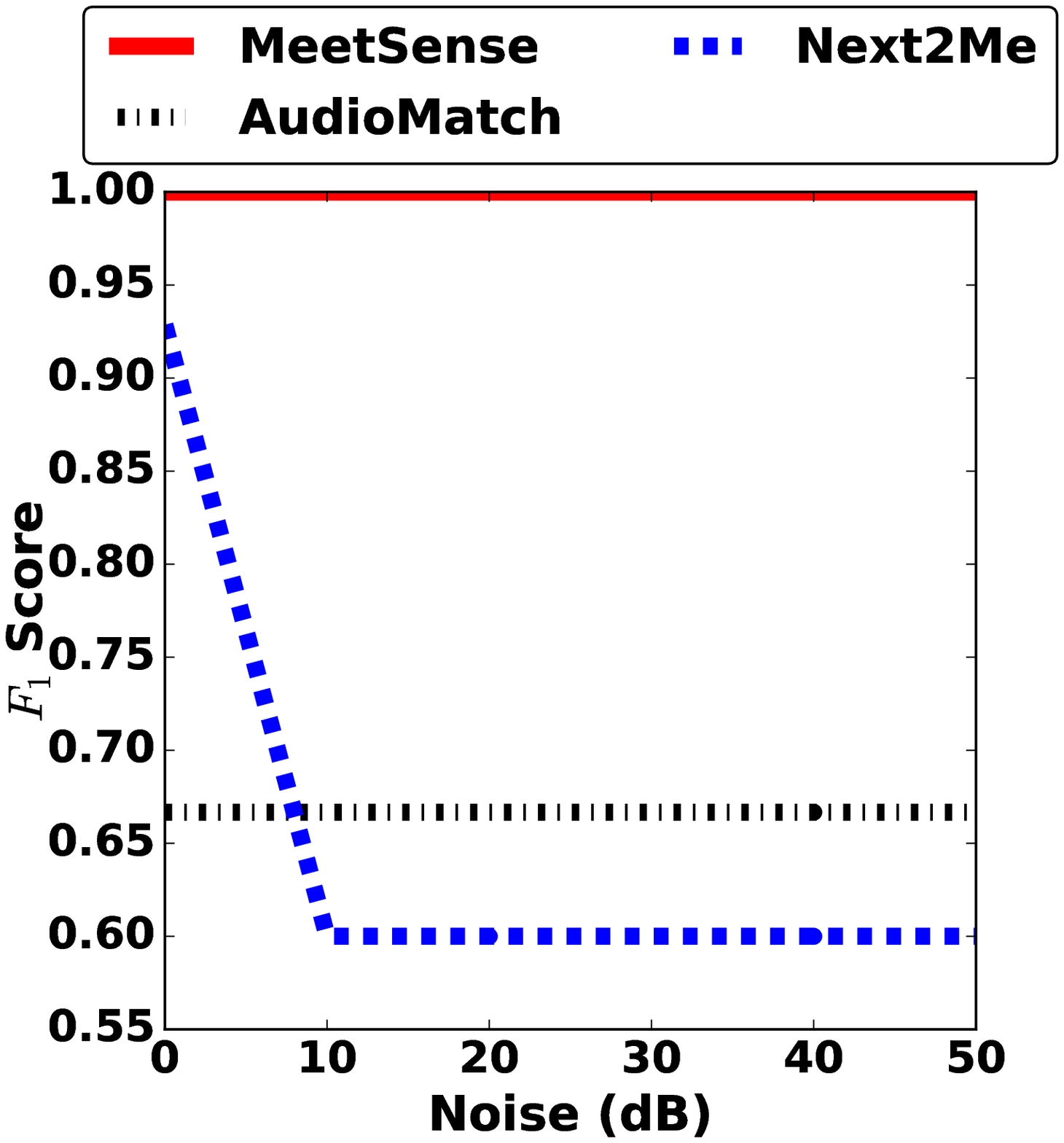}
		\caption{Effect of Noise in\\\textit{MeetSense} and \textit{Next2Me}}
		\label{fig:next2me_noise}
	\end{subfigure}
	\caption{Performance analysis at different environments}
	\label{fig:perAnaly}
\end{figure}

Next, we look into the performance of the various competing methodologies by exploring their internals. From the above experiments, we observe that baselines perform poorly for the scenario $S3$. Therefore, we further study the top $n$ frequency based similarity, thresholding based hamming distance measure, and cepstrum similarity for that scenario. 
%\notesc{Figure \ref{fig:ecdf_compare_s3} is never referenced. This description of is completely messy. What is `far away' and what is `in proximity'? Be very clear while discussing the results. Describe with respect to the scenario. I'll suggest replacing these terms with `same group' and `different groups' throughout the paper. The cross correlation for what? Nothing has been explained properly. Do not assume that the reviewer knows everything and will be able to understand everything just by looking into the figures.} 
From Figure~\ref{fig:ecdf_compare_s3}, we found that audio cross-correlation for `same group' (target subjects are within a group) and `different groups' (target subjects are from different groups) pairs of subjects are more distinct in \textit{MeetSense} as compared with the other two baseline methods. The outdoor environment like the cafeteria (scenario $S3$) are noisy due to the presence of the non-member group voice and noise from the environment. As \textit{Next2Me} considers only the top $6$ frequency components (as per our implementation $n=6$), it unknowingly considers those frequencies, resulting in the similar audio correlation values for `same group' and `different groups'. On the other side, \textit{AudioMatch} compares the logarithmic amplitude of STFT of the audio signal with its neighbouring points to generate $16$-bit fingerprint. Therefore, the $16$-bit fingerprint generation completely relies on the center comparing amplitude value. If the center value is corrupted due to the environmental noise, the entire $16$-bit fingerprint is prone to be corrupted. Those spurious fingerprints are further used for computing the Hamming distance between the pair of subjects, resulting in identical behaviour for the audio correlation values in `same group' and `different groups'. As \textit{MeetSense} considers cepstrum containing the tone information for computing the audio correlation, the correlation values are more close to one for `same group' and close to zero for `different groups'. Therefore, the audio features of \textit{MeetSense} can distinguish this scenario. Consequently, although \textit{Next2Me} and \textit{AudioMatch} fail to separate out the groups based on the audio features, \textit{MeetSense} can correctly differentiate the groups. 

We further evaluate \textit{Next2Me}, \textit{AudioMatch} and \textit{MeetSense} in varying noise environment. As simulating complete random noise is nearly impossible, we generate Gaussian noise at different levels and superimpose the noise with the captured audio signal. Figure \ref{fig:next2me_noise} shows that \textit{MeetSense} is more noise resistant than \textit{Next2Me}, whereas \textit{AudioMatch} is as noise resistant as \textit{MeetSense}. Analogous to noisy environmental scenarios, \textit{Next2Me} performs poorly in the presence of statistically generated Gaussian noise due to the improper selection of top $6$ frequencies. In case of \textit{AudioMatch} the generation of $16$-bit fingerprint causes the drop though it is much less prone to noise as compared to \textit{Next2Me} because of considering the logarithmic amplitude of STFT.
%\notesc{What is the reason for this?}

\begin{figure}[t]
	\centering
	\captionsetup{justification=centering}
	\includegraphics[width=\columnwidth]{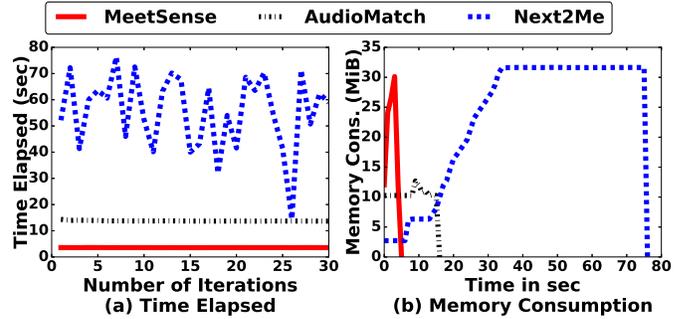}
	\caption{Performance in terms of computational cost for the unsupervised mechanisms}
	\label{fig:perform}
\end{figure}

Next, we compare the three acoustic context detection mechanisms 
%based on unsupervised learning -- \textit{MeetSense}, \textit{AudioMatch}, and \textit{Next2Me} 
in terms of computational resource requirements, as shown in Figure~\ref{fig:perform}. We measure these performance statistics in a standard Linux (Kernel version: 4.4.0) based workstation (Dell Precision Tower 7810) using the \texttt{free} command to obtain the primary memory consumption of the different methodologies. We compute the total execution time and the overall memory consumption during the execution of the three methods. We observe that (i) \textit{MeetSense} takes very less time per iteration during the computation process compared to \textit{Next2Me} and \textit{AudioMatch} (Figure~\ref{fig:perform}a); (ii) the memory consumption for \textit{MeetSense} is less than \textit{Next2Me} (Figure~\ref{fig:perform}b). \textit{MeetSense} enjoys the benefit of lower resource consumption primarily because it computes only cepstrum component for a few segment over the entire interaction time, whereas \textit{Next2Me} uses several windowing operations along with smoothing and FFT computations. \textit{AudioMatch} calculates audio spectrogram using short time fourier transform with a highly overlapping hamming window, causing higher elapsed time than that of \textit{MeetSense}.  
In a nutshell, we observe that \textit{MeetSense} can detect various meeting groups generically and in a device independent way, however, can provide better group detection accuracy with less resource usage compared to the baseline mechanisms.

\subsubsection{\textit{MeetSense} Internals}\label{insight}
In this subsection, we discuss the importance of the modularity value in \textit{MeetSense}, and how the proximity and acoustic context features improve the modularity of the proposed group detection mechanism.

%\subsubsection{Correlation between Modularity and $F_1$-Score}

\begin{figure}[t]
	\centering
	\captionsetup{justification=centering}
	\includegraphics[clip,width=\columnwidth,keepaspectratio]{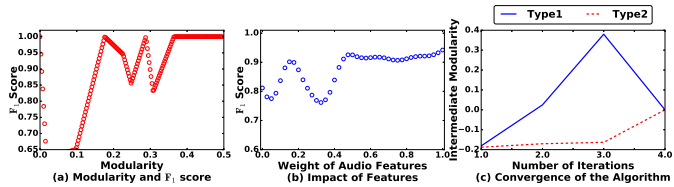}
	\caption{\textit{MeetSense} insights}
	\label{fig:insights}
\end{figure}

We plot the $F_1$-Score with respect to the modularity, as shown in Figure~\ref{fig:insights}a. We observe that the $F_1$-Score converges to $1.0$ when the modularity is more than $0.35$. Hence a group is detected with high accuracy when the cohesiveness is also high. This indicates the importance of modularity index in \textit{MeetSense}. Therefore, the community detection algorithm used in \textit{MeetSense} tries to optimize the modularity in successive iterations.

%\subsubsection{Importance of Proximity and Interaction:}
In this line, Figure~\ref{fig:insights}b highlights the importance of Step 2 of \textit{MeetSense} model, where we plot $F_1$-Score with respect to the weight ($w$) of the audio feature. The figure indicates that the maximum modularity is obtained when both the proximity and the acoustic context attain non-zero weights, indicating that both the features are important for correct detection of meeting groups. However, the importance of acoustic context is more prominent over the proximity feature. 

%\subsubsection{Convergence of the Algorithm:}
Next, we look into the convergence property of \textit{MeetSense}. As mentioned earlier, modularity of a weighted fully connected graph converges to zero when all the nodes form a single large community~\cite{newman2006modularity}. Accordingly, the group detection algorithm converges with two cases of the modularity ($\mathcal{M}$) value -- (a) $\mathcal{M} > 0.0$, when there are multiple groups in the population of subjects ($Type 1$) and (b) $\mathcal{M} \approx 0.0$, when there is a single large group consisting of all the subjects from the population ($Type 2$). Figure~\ref{fig:insights}c plots the change in modularity value with respect to the number of iterations performed in the algorithm, for these two cases. We observe that for a $Type 1$ group corresponding to scenario $S5$, we get the maximum modularity close to $0.4$ with $3$ iterations, whereas for a $Type 2$ group corresponding to scenario $S4$, the modularity starts with a negative value and converges to $zero$ with iteration $4$.

\section{Conclusion} \label{conclusion}
In this paper, we have developed \textit{MeetSense}, a smartphone based light-weight methodology to infer various meeting groups by sensing the acoustic context around the users in proximity. From the pilot study, we have observed that although audio levels captured by a smartphone give a good indication of the acoustic context of the environment, a significant audio pressure from speakers of the nearby groups also gets captured due to the omnidirectional nature of smartphone microphones. We have developed a novel unsupervised methodology to process audio signals to capture the context and used the concept of cohesivity from network science to identify the groups based on context information. The implementation and thorough testing of \textit{MeetSense} shows that it can significantly improve group detection accuracy compared to other baselines, and the method is independent of scenarios or devices used to capture signals. However, our understanding is that \textit{MeetSense} can perform well when the underline groups are sufficiently cohesive; it may fail in the scenarios when multiple groups are overlapped in space, or a group is spatially overlapped with individuals who are not part of that group, for example, small groups in a crowded space. Nevertheless, the proposed methodology has the advantages of device independence, unsupervised modelling and light-weight computation, which can be utilized to develop the wide range of applications that require user group identification and group behaviour analysis. 

\bibliographystyle{IEEEtran}
\bibliography{ref/refernce}

\begin{IEEEbiography} [{\includegraphics[scale=0.1]{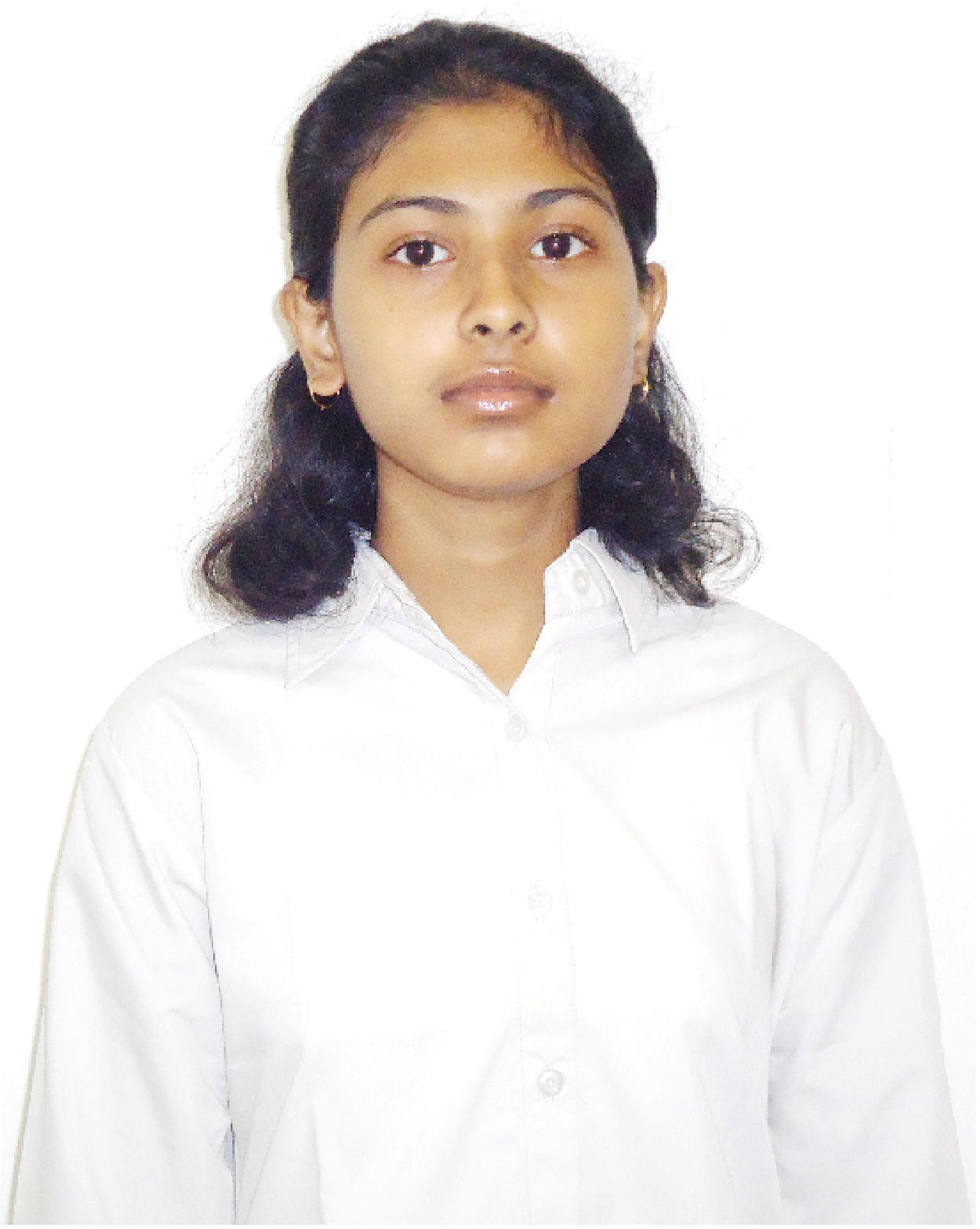}}] {Snigdha Das} is currently pursuing PhD from Department of Computer Science and Engineering, Indian Institute of Technology Kharagpur, India. She received M.S.(by Research) from School of Information Technology, Indian Institute of Technology Kharagpur, India, in 2015. 
	%, and B.Tech. in Information Technology from West Bengal University of Technology, India, in 2010. She also worked as an Associate Software Engineer in CSC for 2 years.
	Her current research interests include mobile systems and ubiquitous computing.
\end{IEEEbiography}

\begin{IEEEbiography} [{\includegraphics[scale=0.5]{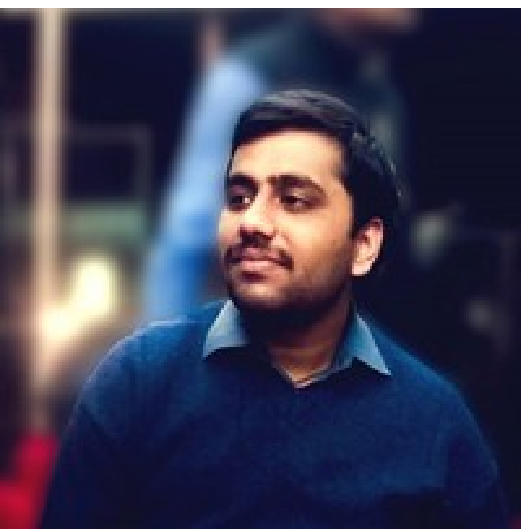}}] {Soumyajit Chatterjee} joined IIT Kharagpur in 2017, as a research scholar (Doctorate Program). He received his M.Tech in Computer Science from Indian Institute of Technology (Indian School of Mines), Dhanbad in the year 2016 and B.E. from University Institute of Technology, University of Burdwan in 2012. He also has an industry experience of one year seven months. Currently, his domain of research is mobile systems and ubiquitous computing.
\end{IEEEbiography}

\begin{IEEEbiography} [{\includegraphics[scale=0.04]{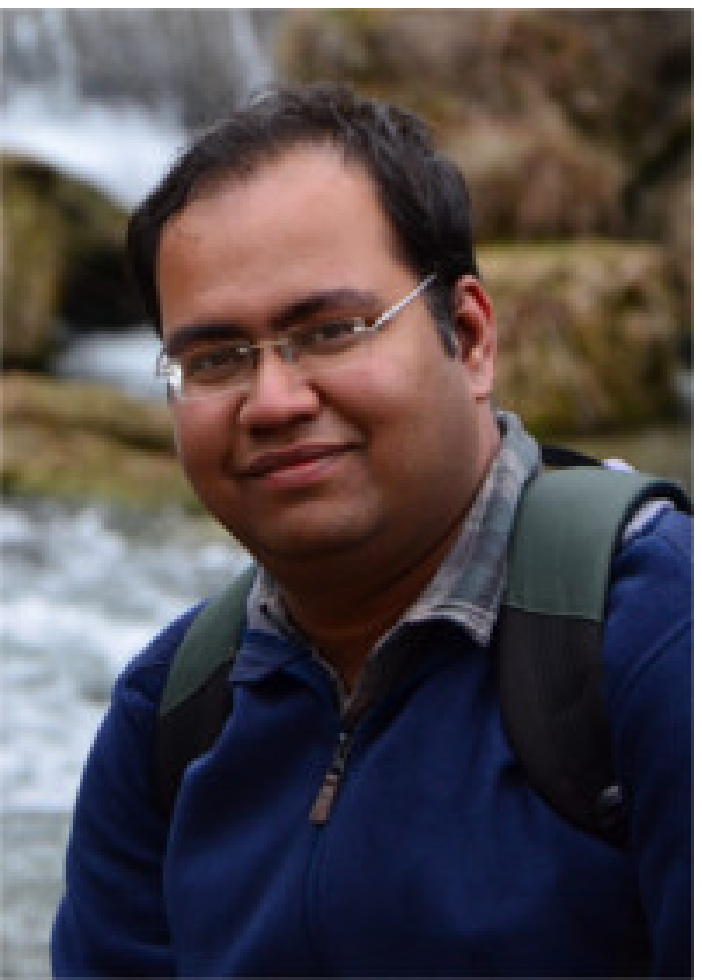}}] {Sandip Chakraborty} received his Ph.D. and M. Tech degrees from the Indian Institute of Technology Guwahati, India, in 2014 and 2011, respectively, and BE in Information Technology from Jadavpur University, Kolkata, India in 2009. Currently, he is an Assistant Professor with the Department of Computer Science and Engineering, Indian Institute of Technology Kharagpur, India. His research interests include computer systems and distributed computing. 
	%He is a Member of the Association for Computing 	Machinery.
\end{IEEEbiography}

\begin{IEEEbiography} [{\includegraphics[scale=0.8]{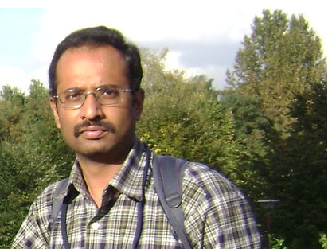}}] {Bivas Mitra} is an Assistant Professor (since April 2013) in the Department of Computer Science \& Engineering at IIT Kharagpur, India. Prior to that, He worked briefly with Samsung Electronics, Noida as a Chief Engineer. He received his Ph.D in Computer Science \& Engineering from IIT Kharagpur, India. He did his first postdoc (May 2010-June 2011) at the French National Centre for Scientific Research (CNRS), Paris, France and second postdoc (July 2011-July 2012) at the Universite catholique de Louvain (UCL), Belgium. His research interests include network science, multilayer network and mobile affective computing. 
	%under the supervision of Prof. Niloy Ganguly and Prof. Sujoy Ghose. 
%	He did his first postdoc (May 2010-June 2011) at the French National Centre for Scientific Research (CNRS), Paris, France and second postdoc (July 2011-July 2012) at the Universite catholique de Louvain (UCL), Belgium.
\end{IEEEbiography}

\vfill 

\end{document}